\documentclass[10pt,twocolumn,twoside]{IEEEtran}
\usepackage{amsfonts,dsfont,amssymb,amsbsy,paralist,bm,ifthen,color}
\usepackage{amsmath}
\usepackage[pdfstartview=FitH,bookmarksnumbered,unicode,bookmarksopen=true]{hyperref}
\usepackage{graphicx}
\usepackage{epstopdf}
\usepackage{multirow}
\usepackage{subfigure,relsize}
\usepackage{mathrsfs}
\usepackage{bm}
\usepackage{stfloats}
\usepackage{url}
\usepackage[noadjust]{cite}
\usepackage{cases,booktabs}
\usepackage{relsize}
\usepackage{algorithm}
\usepackage{algorithmic}

\usepackage{amsthm}  
\makeatletter
\newcounter{subsubsubsection}[subsubsection]

\newcommand\subsubsubsection{\@startsection{subsubsubsection}{4}{\z@}%
  {1.0ex \@plus 1.0ex \@minus .2ex}%
  {0.5ex \@plus .2ex}%
  {\normalfont\normalsize\itshape}}
\newcommand\l@subsubsubsection{\@dottedtocline{4}{7em}{4em}}
\makeatother

\newtheorem{theorem}{Theorem}

\newtheorem{proposition}{Proposition}
\newtheorem{remark}{Remark}
\usepackage[justification=raggedright,singlelinecheck=false]{caption}   
\usepackage[table]{xcolor}
\newcommand\wb{\ensuremath{{\bf w}}}
\newcommand\yb{\ensuremath{{\bf y}}}

\newcommand\Hb{\ensuremath{{\bf H}}}
\newcommand\hb{\ensuremath{{\bf h}}}
\newcommand\Ab{\ensuremath{{\bf A}}}
\newcommand\ab{\ensuremath{{\bm a}}}
\newcommand\Bb{\ensuremath{{\bf B}}}

\newcommand\Cb{\ensuremath{{\bf C}}}

\newcommand\eb{\ensuremath{{\bf e}}}

\newcommand\Gb{\ensuremath{{\bf G}}}

\newcommand\Ib{\ensuremath{{\bf I}}}

\newcommand\pb{\ensuremath{{\bf p}}}
\newcommand\Sb{\ensuremath{{\bf S}}}

\newcommand\Xb{\ensuremath{{\bf X}}}
\newcommand\xb{\ensuremath{{\bf x}}}

\newcommand\Rb{\ensuremath{{\bf R}}}

\newcommand\Qb{\ensuremath{{\bf Q}}}

\newcommand\ub{\ensuremath{{\bf u}}}

\newcommand\Wb{\ensuremath{{\bf W}}}

\newcommand\zb{\ensuremath{{\bm z}}}

\newcommand\Hf{\ensuremath{{\mathsf{H}}}}

\DeclareMathOperator*{\Tr}{Tr}
\graphicspath{{fig/}}

\makeatletter
\renewcommand{\subsubsection}{\@startsection{subsubsection}{3}{\z@}{1.0ex plus 1.0ex minus .2ex}{1.0ex plus .2ex}{\normalfont\normalsize\itshape}}
\makeatother

\begin{document}
%
\title{Optimal Joint Fronthaul Compression and Beamforming Design for Networked ISAC Systems
\\
\thanks{K. Zhang and R. He are with the School of Electronic and Information Engineering, Beijing Jiaotong University, China. (e-mail: kxzhang@bjtu.edu.cn; 	
ruisi.he@bjtu.edu.cn) 

Y. Xu and T.-H. Chang are with School of Science and Engineering, The Chinese University of Hong Kong, Shenzhen 518172, China, and also with the Shenzhen Research Institute of Big Data, Shenzhen 518172, China (e-mail: xuyanqing@cuhk.edu.cn; tsunghui.chang@ieee.org).

C. Shen is with the Shenzhen Research Institute of
Big Data, Shenzhen 518172, China (e-mail: chaoshen@sribd.cn).

}}
%
%
%

\author{Kexin~Zhang,~\IEEEmembership{Student Member,~IEEE,}
        Yanqing~Xu,~\IEEEmembership{Member,~IEEE,}
        Ruisi~He,~\IEEEmembership{Senior~Member,~IEEE,}
        Chao~Shen,~\IEEEmembership{Member,~IEEE,} 
        and~Tsung-Hui~Chang,~\IEEEmembership{Fellow,~IEEE}}

%
%


\maketitle

\begin{abstract}
This study investigates a networked integrated sensing and communication (ISAC) system, where multiple base stations (BSs), connected to a central processor (CP) via capacity-limited fronthaul links, cooperatively serve communication users while simultaneously sensing a target. The primary objective is to minimize the total transmit power while meeting the signal-to-interference-plus-noise ratio (SINR) requirements for communication and sensing under fronthaul capacity constraints, resulting in a joint fronthaul compression and beamforming design (J-FCBD) problem. We demonstrate that the optimal fronthaul compression variables can be determined in closed form alongside the beamformers, a novel finding in this field. Leveraging this insight, we show that the remaining beamforming design problem can be solved globally using the semidefinite relaxation (SDR) technique, albeit with considerable complexity. 
Furthermore, the tightness of its SDR reveals { zero} duality gap between the considered problem and its Lagrangian dual. Building on this duality result, we exploit the novel UL-DL duality within the ISAC framework to develop an efficient primal-dual (PD)-based algorithm. The algorithm alternates between solving beamforming with a fixed dual variable via fixed-point iteration and updating dual variable via bisection, ensuring global optimality and achieving high efficiency due to the computationally inexpensive iterations. Numerical results confirm the global optimality, effectiveness, and efficiency of the proposed PD-based algorithm.


\end{abstract}


\IEEEpeerreviewmaketitle

\section{Introduction}

\IEEEPARstart{I}{ntegrated} sensing and communications (ISAC) is emerging as a cornerstone technology for future wireless networks \cite{liu2020joint, cui2021integrating}. 
To support a range of intelligent applications, e.g., smart cities and autonomous driving, this innovative approach enables concurrent target sensing and data transmission \cite{liu2020joint, cui2021integrating}.
The ability to merge these two functions into a single framework opens up new avenues for optimizing the performance and functionality of next-generation wireless systems. By sharing spectral resources and reusing hardware architectures, ISAC not only enhances spectral efficiency but also significantly reduces implementation costs. Numerous works have studied this topic, primarily concentrating on beamforming design, echo signal processing, and resource allocation for a single-cell ISAC \cite{liu2021cramer, ren2023fundamental, xiao2022waveform}.

Recently, networked ISAC { \cite{cai2024sensing, zhu2023information, ji2023networking, li2024networked, huang2022coordinated, cheng2024optimal, liu2024joint,zhang2022active, zhang2024beamforming}} (or cell-free ISAC \cite{behdad2022power, demirhan2023cell, dong2024cell}, perceptive mobile networks \cite{zhang2020perceptive, rahman2019framework, lei2023coll}), which integrates sensing capabilities into current wireless networks, further capitalizes on the potential of ISAC compared to conventional single-cell ISAC and attracts tremendous research momentum.
Leveraging the cloud radio access network (C-RAN) architecture, a networked ISAC system consisting of multiple base stations (BSs) and a central processor (CP) benefits from multiple BSs coordination and joint signal processing \cite{rao2014distributed, zhou2016fronthaul}. 
In terms of communication, networked ISAC facilitates coordinated multi-point (CoMP) data transmission, effectively reducing inter-user interference and significantly boosting data rates \cite{gesbert2010multi, xiang2013coor}. On the sensing front, networked ISAC excels in distributed radar sensing, markedly expanding coverage areas, providing diverse sensing angles, and enhancing sensing accuracy \cite{liang2024compress, liang2011design, haimovich2007mimo, he2010target}.
Additionally, it offers a pragmatic solution to the full-duplex challenge in single-cell ISAC systems by designating some BSs as dedicated sensing receivers, thereby ensuring effective interference control between reflected signals and downlink (DL) ISAC signals.
These benefits collectively demonstrate the transformative potential of networked ISAC in modern cellular networks.

Despite the great potential of networked ISAC, the inherently limited fronthaul link capacity in the C-RAN poses a significant challenge {to the} system design. This is because the cooperation between BSs leads to substantial information exchange between the CP and BSs, exacerbating the burden on these fronthaul links {\cite{xu2024distributed, patil2018hybrid, yu2016energy, kim2019joint}}.
To tackle the above issue, the compression techniques in fronthaul transmission have been proposed to reduce the number of transmission bits \cite{yu2016energy,kim2019joint,liu2021uplink, fan2023qos}. Efficient fronthaul compression plays a critical role in maintaining the high-precision sensing and reliable communication expected from networked ISAC systems. Therefore, transmit beamformers should be jointly designed along with the utilization of fronthaul compression. In this work, we investigate the joint fronthaul compression and beamforming design (J-FCBD) for networked ISAC systems. Along this direction, only a few works have been proposed  \cite{dong2024cell, zhang2022active, zhang2024beamforming}. 
Compared to these works, we further investigate the {low-complexity} algorithm design, exploiting the primal-dual (PD) optimization and uplink (UL)-DL duality in networked ISAC. Although the PD-based algorithm for J-FCBD has been exploited in communication system \cite{yu2016energy,kim2019joint,liu2021uplink, fan2023qos}, it cannot be directly applied to network ISAC because the sensing requirements make the problem more complicated and bring new challenges.
This paper aims to bridge this gap by exploiting the duality result in the J-FCBD problem within networked ISAC, unveiling its special solution structure and harnessing it to devise low-complexity algorithms

\subsection{Related Works}
In the literature, several works have investigated networked ISAC, such as \cite{zhang2020perceptive, huang2022coordinated, cheng2024optimal, liu2024joint}.
For instance, \cite{zhang2020perceptive} presented the system architecture of networked ISAC by referring to distributed antenna systems, in particular, C-RAN.
Leveraging C-RAN architecture, \cite{huang2022coordinated} examined a multi-cell networked ISAC system with single-antenna BSs. In this system, BSs jointly optimized their transmit power control to minimize total transmit power while meeting the signal-to-interference-plus-noise ratio (SINR) constraints for communication users and ensuring the estimation accuracy for target localization.
Additionally, \cite{cheng2024optimal} further explored a multi-antenna networked ISAC system, maximizing the detection probability under communication SINR requirements and power constraints via jointly optimizing the communication and dedicated sensing beamforming.
Similarly, \cite{liu2024joint} focused on the joint design of transmit beamformer and receive filter for multi-antenna networked ISAC to maximize sensing SINR under the communication SINR constraints and power budget.
However, these studies \cite{huang2022coordinated, cheng2024optimal, liu2024joint} assumed ideal fronthaul capacity, which { does not hold true} in practical scenarios where { the} fronthaul capacity is limited. 
Overlooking the impact of fronthaul capacity limitation leads to {overestimated performance}, resource wastage, and operational inefficiencies. 
{ For the networked ISAC, integrating the fronthaul compression with beamforming design becomes crucial since it impacts on both the communication and sensing performance.
Therefore,} the J-FCBD framework is essential to addressing these practical constraints and ensuring optimal performance of networked ISAC systems.


While the J-FCBD has been explored extensively {in communication-only wireless networks} \cite{yu2016energy,kim2019joint,liu2021uplink, fan2023qos}, these approaches cannot be directly applied to networked ISAC due to the sensing requirements { introduced by} ISAC. This adds complexity and poses significant challenges to the J-FCBD algorithm design.
In the literature of the J-FCBD for networked ISAC, research remains limited \cite{dong2024cell, zhang2022active, zhang2024beamforming}. 
For instance, \cite{dong2024cell} considered a zero-forcing (ZF) transmit beamforming-based networked ISAC, in which the CP jointly optimized power allocation and fronthaul compression to maximize the sensing SINR while ensuring the minimum required communication SINR and adhering fronthaul capacity limitation. However, the J-FCBD algorithm using ZF beamforming strategy led to performance loss. 
Furthermore, a few works \cite{zhang2022active, zhang2024beamforming} focused on the J-FCBD for intelligent reflecting surface (IRS)-assisted networked ISAC scenarios.  In \cite{zhang2022active}, active IRS was deployed to amplify and reflect the joint communication and sensing signals from BSs. The J-FCBD problem was formulated to form the desired reflective beam pattern towards the sensing targets while satisfying communication SINR requirements and fronthaul capacity limits. 
In \cite{zhang2024beamforming}, J-FCBD was exploited for wideband IRS-assisted orthogonal frequency division multiplexing (OFDM)-ISAC system. The goal was to minimize the beampattern mismatch error subject to communication user rate constraints and fronthaul capacity constraints. Note that the J-FCBD algorithm in \cite{zhang2022active, zhang2024beamforming} employed an alternating approach, using semidefinite relaxation (SDR) and Taylor expansion technique. These alternating subproblems are solvable by off-the-shelf solvers like CVX \cite{grant2014cvx} but are typically of considerable computational complexity.


\subsection{Contributions}
    In practical systems, fronthaul capacity limitations hinder achieving excellent networked ISAC performance due to the substantial burden on the fronthaul link caused by BS cooperation. In light of this, we investigate a networked ISAC system where a CP is connected to multiple BSs via capacity-limited fronthaul links to perform cooperative communication and sensing.
    By explicitly incorporating the fronthaul capacity limitation into the networked ISAC framework, we formulate the J-FCBD problem of optimizing system performance under realistic constraints. Specifically, the J-FCBD problem aims to minimize the total transmit power while satisfying the SINR requirements of both communication and sensing, subject to both DL and UL fronthaul compression. Due to the non-convexity and the large number of constraints, the problem is complex and challenging to solve.

    To solve the complex J-FCBD problem, we first demonstrate that the optimal UL and DL fronthaul compression variables can be determined in closed form alongside the beamformers. Specifically,  it is found that the SDR technique is indispensable to drive the DL one. Subsequently, we devise an SDR-based algorithm to solve the remaining beamforming design problem and establish the tightness of the SDR problem, implying its global optimality and revealing the hidden convexity of the problem. 
    This insight enables us to exploit PD optimization for effectively handling the problem in a low-complexity fashion. 
    An interesting result is that the UL-DL duality, a pivotal result in classical communication theory { \cite{viswanath2003sum, liu2021uplink}}, still holds in the networked ISAC setting. This remarkable extension of the duality principle to ISAC scenarios underscores the robustness and versatility of the duality concept. Based on this insight, we develop a low-complexity algorithm that alternates between solving the beamforming problem with respect to the dual variable via the fixed-point method and updating the dual variable via bisection. 
    This algorithm is guaranteed to find the global solution and is highly efficient due to the computationally inexpensive updates in both the fixed-point and bisection iterations.

    Extensive simulations are conducted to validate the effectiveness and efficiency of the proposed algorithms. Simulation results show that both the SDR-based algorithm and the PD-based algorithm achieve a significant power saving compared to the separated design benchmark. Moreover, the computational time of PD-based algorithm is remarkably reduced compared to that of the SDR-based algorithm without performance loss in power saving.
    Additionally, the simulation results show that the fronthaul capacity has a significant impact on the system performance, confirming the necessity of considering the limited fronthaul capacity in networked ISAC designs.

    \textbf{Synopsis}: Section \ref{section_system} introduces the networked ISAC system with limited fronthaul capacity and gives the J-FCBD problem formulation. The SDR-based algorithm is proposed in Section \ref{section_SDR}. Section \ref{section_PD} presents a {low-complexity} algorithm based on PD optimization.  Section \ref{section_simulation} evaluates the performances of the proposed algorithms by simulations. Finally, the conclusion is drawn in Section \ref{section_conclusion}.

    \textbf{Notations}: Column vectors and matrices are denoted by boldfaced lowercase and uppercase letters, e.g., $\xb$ and $\Xb$. $\mathbb{R}^{n \times n}$ and $\mathbb{C}^{n \times n}$ stand for the sets of $n$-dimensional real and complex matrices, respectively. The superscripts $(\cdot)^{\top}$ and $(\cdot)^{\Hf}$ describe the transpose and Hermitian operations, respectively. $\Tr(\Xb)$ represents the trace of $\Xb$. $\operatorname{diag}\left(x_1, \dots, x_N\right)$ returns a diagonal matrix with $x_1, \dots, x_N$ are on the principle diagonal of the matrix. 
    $\|\xb\|^2$ denotes {the squared} Euclidean norm of vector $\xb$. $\Xb^{-1}$ denotes the inverse operation.

\section{System Model and Problem Formulation}\label{section_system}
\begin{figure}[!t]   \centering
    \includegraphics[width=8.6 cm]{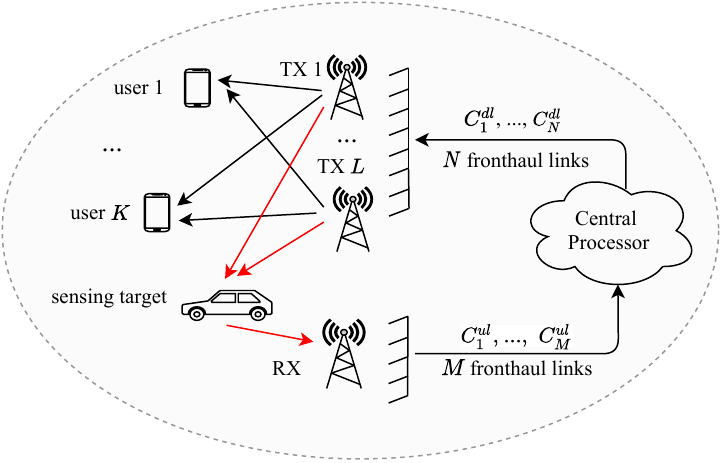}
    \caption{\small An illustration of the networked ISAC system with capacity-limited fronthaul links.}
    \label{fig: system model}
\end{figure}

Consider a networked ISAC system where a CP coordinates {$L$} ISAC transmitters (TXs) and a sensing receiver (RX) to serve $K$ single-antenna { communication} users and meanwhile sense one point target, as shown in Fig. \ref{fig: system model}. Assume that the number of transmit antennas is $N$ and the number of receive antennas is $M$. Each TX is equipped with $N_t \triangleq \frac{N}{L}$ antennas.
To perform joint transmit beamforming and target sensing, the TXs and the RX communicate with the CP via the fronthaul links \cite{patil2018hybrid, yu2016energy,kim2019joint}. In this paper, for facilitating practical applications, we assume that the fronthaul bandwidth is limited and only the compressed signals with limited number of transmission bits are exchanged between the CP and the TXs/RX.

\subsection{Communication Model}
Let $s_{k}$ be the signal for user $k$, and $\wb_{l,k}$ be the associated beamforming vector generated at the CP and conveyed to TX $l$. For ease of presentation, we define a super-vector $\wb_k = [\wb_{1,k}^\top, \dots, \wb_{L,k}^\top]^\top \in \mathbb{C}^{N\times 1}$ to collect all the beamforming vectors for serving user $k$.
The signal to be transformed from the CP to TXs is given by
\begin{align}
    \xb = \sum_{k=1}^K \wb_{k} s_k,
\end{align}
where $\xb = [x_1, \dots, x_N]^\top$ with $x_n$ signifying the transmit signal for the $n$-th transmit antenna.
Rather than delivering the complete $\xb$ from the CP to TXs, a compressed signal is transmitted by using the compress-then-transmit strategy for fronthaul bandwidth saving.
Here, the impact of compression is modeled by assuming that the fronthaul link is a Gaussian test channel \cite{patil2018hybrid,yu2016energy,kim2019joint}. 
Denote the DL compression noise introduced in the compression process as $\eb^{dl} \in \mathbb{C}^{N\times 1}$, which is modeled as an independent Gaussian random variable, i.e., $\eb^{dl} \sim \mathcal{CN}(\mathbf{0}, \Qb_{dl})$, and $\Qb_{dl} = \operatorname{diag}\left(q_{1}^{dl}, \dots, q_{N}^{dl}\right)$ denotes its covariance matrix. 
Accordingly, the received compressed signal for the TXs is given by 
\begin{align}\label{compress}
\bar \xb = \xb + \eb^{dl},
\end{align}
where $\bar \xb=[\bar x_1, \dots, \bar x_N]^\top$ with $\bar x_n$ signifying the compressed transmit signal for the $n$-th transmit antenna. 

Based on the above model, the DL fronthaul rate for transmitting $\bar{x}_n$ is given by \cite{liu2021uplink, kim2019joint}
\begin{align}
R_{n}^{dl} &= I(x_n; \bar{x}_n) \notag\\
&= \log \left(1+\frac{\sum_{k=1}^K \wb_k^\Hf \eb_n \eb_n^\top\wb_k}{q_n^{dl}}\right), \forall n \in \mathcal{N},
\end{align}
where $\eb_n$ is a unit vector with only the $n$-th element being $1$, and $\mathcal{N} \triangleq \{1, 2, \dots, N\}$ denotes the set of transmit antennas.

After receiving the compressed signal, the TXs transmit it to the $K$ users. 
Denote $\hb_{k}\in \mathbb{C}^{N \times 1}$ as the channel between TXs and user $k$, and $v_k \sim \mathcal{CN}(0,\sigma^2_v)$ as the additive white Gaussian noise (AWGN) at user $k$. The received signal at user $k$ can be expressed as 
\begin{align}
    y_k = {\hb_k}^\Hf\bar\xb  + v_{k}, \forall k \in \mathcal{K},
\end{align}
where $\mathcal{K} \triangleq \{1, 2, \dots, K\}$ denotes the set of users.
Therefore, the achievable SINR for the communication of user $k$ can be written as 
\begin{align}\label{SINR}
    \operatorname{SINR}^c_{k}
    &= \frac{|\hb_{k}^\Hf \wb_{k}|^2}{ \sum_{i \neq k} |\hb_{k}^\Hf \wb_i|^2  + \hb_{k}^\Hf \Qb_{dl}\hb_{k} + \sigma^2_v}, \forall k \in \mathcal{K}.
\end{align}

\subsection{Sensing Model}
The target sensing is based on the received reflected signal at RX, as shown in Fig. \ref{fig: system model}. Here, we assume that the TXs-target-RX channel is dominated by the line-of-sight (LoS) path between TXs/RX and the target, while the other weak non-LoS paths are ignored \cite{liu2020radar, huang2022coordinated}.
In particular, the TXs-target-RX channel can be modeled by
\begin{align} \label{sensingchannel}
    \mathbf{G} = \ab_r\ab_t^\Hf \boldsymbol{\Sigma}_g.
\end{align}
where $\boldsymbol{\Sigma}_g \triangleq \operatorname{diag}(g_1, \dots, g_L)\otimes \mathbf{I}_{N_t}$, $g_l$ is the coefficient of the signal propagation path of {the $l$-th TX-target-RX channel}, $\ab_t \triangleq [\ab_{t,1}^\top, \dots, \ab_{t,L}^\top]^\top$ with  $\ab_{t,l}$ denoting the transmit steering vector for { the} $l$-th TX, and $\ab_r$ is the receive steering vector \cite{liu2020radar}, 
\begin{align}
\ab_{t,l} & =\sqrt{\frac{1}{N_t}}\left[1, e^{-j \pi \cos \theta_{t,l}}, \ldots, e^{-j \pi\left(N_t-1\right) \cos \theta_{t,l}}\right]^\top, \forall l \in \mathcal{L},\notag\\
\ab_r & =\sqrt{\frac{1}{M}}\left[1, e^{-j \pi \cos \theta_r}, \ldots, e^{-j \pi\left(M-1\right) \cos \theta_r}\right]^\top.
\end{align}
Here $\theta_{t,l}$ is the direction of departure (DOD) of the $l$-th TX, $\theta_r$ is the direction of arrival (DOA), and $\mathcal{L} \triangleq \{1, \dots, L\}$ denotes the set of TXs. 

Based on the above channel model, the received reflected signal at RX is 
\begin{align}\label{RX}
\yb &= \mathbf{G} \bar \xb + \zb,
\end{align}
where $\zb \sim \mathcal{CN}(0,\sigma^2_z\Ib)$ is AWGN.

After receiving the reflected signal, the RX transmits it to the CP via UL fronthaul links for sensing signal processing.
Similar to the DL fronthaul transmission, the compress-then-transmit strategy is utilized, and the UL compression noise is modeled as $\eb^{ul} \sim \mathcal{CN}(0, \Qb_{ul})$ with covariance matrix $\Qb_{ul} = \operatorname{diag}(q^{ul}_1, \dots, q^{ul}_{M})$.
The compressed signal of $\yb$ can be expressed as
\begin{align} \label{bar{r}}
\bar{\yb} = \yb + \eb^{ul}.
\end{align}
Let $y_m$ and $\bar{y}_m$ signify the received signal and compressed signal for the $m$-th RX antenna, respectively. The fronthaul rate for transmitting $\bar{y}_m$ can be expressed as \cite{liu2021uplink, kim2019joint}
\begin{align}\label{r_ul}
    R_m^{ul} & = I(y_m; \bar{y}_m) \notag\\
    &= \log\left(1+ \frac{ \ab_{t}^\Hf \boldsymbol{\Sigma}_g \Rb \boldsymbol{\Sigma}_g^\Hf \ab_{t} + M\sigma_z^2}{M{q}_m^{ul}}\right), \forall m \in \mathcal{M},
\end{align}
where $\mathcal{M} \triangleq \{1, 2, \dots, M\}$ denotes the set of RX antennas, and the matrix $\Rb$ is defined as
\begin{align}
\Rb \triangleq \sum_{k=1}^K \wb_k\wb_k^\Hf + \Qb_{dl}.
\end{align}

With a receive beamforming vector $\ub \in \mathbb{C}^{M}$, the received compressed signal at CP is given by
\begin{align}\label{haty}
    \hat{y} = \ub^\Hf \bar{\yb}.
\end{align}
Based on \eqref{haty}, the  sensing SINR can be calculated as
\begin{align} \label{SINR_S}
    \operatorname{SINR}^s = \frac{\ub^\Hf \Gb \Rb \Gb^\Hf \ub } {\ub^\Hf (\Qb_{ul} + \sigma^2_z \Ib)\ub},
\end{align}
which is used to measure the target sensing performance \cite{gezici2005localization}.

The optimal $\ub^*$ to maximize the sensing SINR can be derived by solving the equivalent minimum variance distortionless response (MVDR) problem as in \cite{cox1987robust}, which is given by
\begin{align} \label{v}
    \ub^* =(\Qb_{ul} + \sigma^2_z \Ib)^{-1}\ab_r. 
\end{align}
By substituting $\ub^*$ into \eqref{SINR_S}, the sensing SINR can be rewritten as
\begin{align}
    \operatorname{SINR}^s = \left(\ab_t^\Hf  \boldsymbol{\Sigma}_g \Rb  \boldsymbol{\Sigma}_g^\Hf \ab_t \right) \ab_r^\Hf (\Qb_{ul} + \sigma^2_z \Ib)^{-1}\ab_r.
\end{align}

\subsection{Problem Formulation}

Based on the above model, we formulate the J-FCBD problem, which aims to minimize the total transmit power while guaranteeing the SINR requirement of sensing and communication, and fronthaul capacity constraints,
\begin{subequations} \label{P1}
    \begin{align}
 \text{(P1)}:\  &  \min _{\{\wb_k\}, \{q_n^{dl}\}, \{q_m^{ul}\}} \   \Tr(\Rb) \\
     \text { s.t. } &
        \log \left(1+\frac{\sum_{k=1}^K \wb_k^\Hf \eb_n \eb_n^\top\wb_k}{q_n^{dl}}\right) \leq C^{dl}, \forall n \in \mathcal{N}, \label{P1-1}\\
        & \frac{|\hb_{k}^\Hf \wb_{k}|^2}{ \sum\limits_{i \neq k} |\hb_{k}^\Hf \wb_i|^2  + \hb_{k}^\Hf \Qb_{dl}\hb_{k} + \sigma^2_v} \geq \Gamma_k, \forall k \in \mathcal{K}, \label{P1-2}\\
        &  \log\left(1+ \frac{\ab_{t}^\Hf \boldsymbol{\Sigma}_g \Rb \boldsymbol{\Sigma}_g^\Hf \ab_{t} + M\sigma_z^2}{M{q}_m^{ul}}\right) \leq C^{ul}, \forall m \in \mathcal{M},\label{P1-3} \\
        & \left( \ab_t^\Hf  \boldsymbol{\Sigma}_g \Rb  \boldsymbol{\Sigma}_g^\Hf \ab_t \right) \ab_r^\Hf (\Qb_{ul} + \sigma^2_z \Ib)^{-1}\ab_r \geq \Gamma_s, \label{P1-4}\\
        & \ q_n^{dl} \geq 0,\forall n \in \mathcal{N}, \ q^{ul}_m \geq 0, \forall m \in \mathcal{M},\label{P1-5}
    \end{align}
\end{subequations}
where $C^{dl}$ and $C^{ul}$ are the capacities of the DL fronthaul links and UL fronthaul links, respectively, $\Gamma_k$ is the required communication SINR for the $k$-th user, and $\Gamma_s$ is the required SINR for sensing.

Several works have studied problems similar to (P1), such as \cite{liu2021uplink,cheng2024optimal,dong2024cell}. However, (P1) presents greater challenges, making existing algorithms inapplicable. 
Specifically, the work in \cite{liu2021uplink} focused on the J-FCBD problem for DL communication systems, {whereas problem (P1) is evidently more difficult to handle as the UL constraints \eqref{P1-3}-\eqref{P1-4} are coupled with the DL constraints \eqref{P1-1}-\eqref{P1-2}}. 
{In addition, the extra compression variables in (P1) significantly increase its complexity compared to the study in \cite{cheng2024optimal} which assumes the ideal fronthaul capacity.}
Finally, although \cite{dong2024cell} considered the fronthaul compression and beamforming design problem for ISAC systems, they are not jointly optimized. In particular, a suboptimal zero-forcing precoding scheme is considered, and the compression is determined by a codebook-based scheme, {in contrast to (P1) which allows for} more flexible beamforming and compression designs.

Interestingly, as will be shown in the following proposition that the optimal UL fronthaul noise level, $\{q_m^{ul}\}$, of problem (P1) actually admits closed-form solution. This finding helps simplify problem (P1) into a more tractable form.
\begin{proposition}\label{lemmaul}
    Suppose that problem (P1) is feasible. Then, we have that the optimal $\{{q}_m^{ul,*}\}$ always makes constraint \eqref{P1-3} hold with equality, i.e.,
\begin{align} \label{qul}
    {q}_m^{ul,*} = \frac{\beta}{M} \ab_{t}^\Hf \boldsymbol{\Sigma}_g \Rb^* \boldsymbol{\Sigma}_g^\Hf \ab_{t} + \beta\sigma_z^2, \forall m \in \mathcal{M},
\end{align} 
 where $\beta = \frac{1}{2^{C^{ul}}-1} $ is a constant.
 \\
 {\bf Proof:} The proof is relegated to Appendix \ref{A}. \hfill $\blacksquare$
\end{proposition}

By substituting \eqref{qul} into (P1), it {boils} down to
\begin{subequations} \label{P2}
    \begin{align}
  \text{(P2)}:\    & \min _{\{\wb_k\}, \{q_n^{dl}\}} \   \Tr\bigg(\sum_{k=1}^K \wb_k\wb_k^\Hf + \Qb_{dl}\bigg)\\
    \text { s.t. } 
    & \alpha \sum_{k=1}^K \eb_n^\top\wb_k\wb_k^\Hf\eb_n \leq  q_n^{dl}, \forall n \in \mathcal{N}, \label{P2-1}\\
    & \frac{|\hb_{k}^\Hf \wb_{k}|^2}{ \sum\limits_{i \neq k} |\hb_{k}^\Hf \wb_i|^2  + \hb_{k}^\Hf \Qb_{dl}\hb_{k} + \sigma^2_v} \geq \Gamma_k, \forall k \in \mathcal{K}, \label{P2-2}\\
    &\ab_t^\Hf \boldsymbol{\Sigma}_g \bigg(\sum_{k=1}^K \wb_k\wb_k^\Hf + \Qb_{dl}\bigg) \boldsymbol{\Sigma}_g^\Hf \ab_t \geq \widetilde{\Gamma}_s, \label{P2-3} \\
    & \ q_n^{dl} \geq 0,\forall n \in \mathcal{N},
    \end{align}
\end{subequations}
where $\alpha \triangleq \frac{1}{2^{C^{dl}} -1}$ and $\widetilde{\Gamma}_s \triangleq \frac{M\Gamma_s(1+\beta)\sigma_z^2}{M -\Gamma_s\beta}$. 

Compared to (P1), (P2) has a more concise form due to the absence of UL compression variable designs. However, it remains challenging to solve because of  the coupling between beamformers and DL compression variables in \eqref{P2-2}, and the non-convexity of \eqref{P2-3}. In the following two sections, we present two efficient algorithms to solve (P2).


\vspace{1em}

\section{Optimal Solution via SDR Method}\label{section_SDR}
As seen in (P2), the nonconvexity arises from the quadratic terms related to the beamformers. To address this challenge, we leverage the popular SDR method to approximate these quadratic terms \cite{luo2010semidefinite}. Interestingly, as will be shown later in this section, we can actually obtain globally optimal solutions for (P1) using the SDR method, i.e., the SDR tightness is guaranteed.


{ Define} the matrices $\Wb_k \triangleq \wb_k\wb_k^\Hf$ with $\operatorname{rank}(\Wb_k) = 1, \forall k \in \mathcal{K}$. The SDR method  ignores the rank constraints and solves the following approximated problem:
\begin{subequations} \label{SDR2}
    \begin{align}
 &\text{(SDR1)}: \     \min _{\{\Wb_k\}, \{q_n^{dl}\}} \   \Tr\left(\sum_{k=1}^K \Wb_k +\Qb_{dl}\right)\label{SDR2-0}\\
    &\text { s.t. } 
         \alpha \sum_{k=1}^K \eb_n^\top\Wb_k\eb_n \leq  q_n^{dl}, \forall n \in \mathcal{N}, \label{SDR2-1}\\
        & \qquad \frac{\hb_{k}^\Hf \Wb_{k}\hb_{k} }{  \hb_{k}^\Hf \left( \sum_{i \neq k}\Wb_i  + \Qb_{dl}\right)\hb_{k} + \sigma^2_v} \geq \Gamma_k, \forall k \in \mathcal{K}, \label{SDR2-2}\\
        &  \qquad \ab_t^\Hf \boldsymbol{\Sigma}_g  \left(\sum_{k=1}^K \Wb_k +\Qb_{dl}\right)\boldsymbol{\Sigma}_g^\Hf \ab_t \geq \widetilde{\Gamma}_s, \label{SDR2-4}\\
        & \qquad \ q_n^{dl} \geq 0,\forall n \in \mathcal{N}, \Wb_k \succeq \mathbf{0}, \forall k \in \mathcal{K}.\label{SDR2-5}
    \end{align}
\end{subequations}
Problem (SDR1) is convex. {A key observation which paves the optimality of SDR is that the optimal DL compression variable $\{q_n^{dl,*}\}$ of (SDR1) has a closed expression, as shown in the following proposition.}

\begin{proposition}\label{lemmadl}
Suppose that problem (SDR1) is feasible. Then, problem (SDR1) always has an optimal solution such that constraint \eqref{SDR2-1} holds with equality, i.e.,
\begin{align} \label{qdlSDR}
    q_{n}^{dl,*} = \alpha \sum_{k=1}^K \eb_n^\top\Wb_k^*\eb_n, \forall n \in \mathcal{N}.
\end{align}
\end{proposition}

{\bf Proof:} The proof is relegated to Appendix \ref{B}. \hfill $\blacksquare$

{
\begin{remark}
We should emphasize that the result in Proposition 2, i.e., the DL noise variable can be optimally represented by the beamforming variables, cannot be directly { inferred} from the original problem (P2). This is because, in (P2), 
{it is not possible to construct a contractive solution (e.g., \eqref{newsolution1} in Appendix \ref{B}) when \eqref{qdlSDR} does not  hold true, due to the non-linear structure of (P2). This indicates the essence of applying SDR since (SDR1) is linear and \eqref{newsolution1} is feasible.}
\end{remark}
}

Based on Proposition \ref{lemmadl}, solving (SDR1) can be further simplified by equivalently solving the following problem:
\begin{subequations} \label{P4}
  \begin{align}
    & \text{(SDR2):}\    \min _{\{\Wb_k\}} \ (1+\alpha) \sum_{k=1}^K \Tr\left(\Wb_k\right)\\
    &\text { s.t. } 
         \widetilde{\Gamma}_k \Tr\left(\Wb_{k} \Hb_{k}\right) \geq  \sum_{i =1}^K \Tr\left(\Wb_{i} \Ab_k\right) + \sigma^2_v, \forall k \in \mathcal{K}, \label{P4-2}\\
        & \qquad \sum_{k =1}^K \Tr\left( \Wb_{k} \Bb\right) \geq \widetilde{\Gamma}_s, \label{P4-4}\\
        & \qquad \Wb_k \succeq \mathbf{0}, \forall k \in \mathcal{K},
    \end{align}
\end{subequations}
where $\widetilde{\Gamma}_k \triangleq 1 + \frac{1}{\Gamma_k}$, $\Hb_{k} \triangleq \hb_{k}\hb^\Hf_{k}$, $\Ab_k \triangleq  \Hb_{k} + \alpha \operatorname{diag}(\Hb_{k})$, and $\Bb \triangleq \boldsymbol{\Sigma}_g^\Hf \left(\ab_t \ab_t^\Hf +\frac{\alpha}{N_t} \Ib \right)\boldsymbol{\Sigma}_g$. {
Interestingly, the following theorem shows that (SDR2) has a rank-one solution for $\{\Wb_k^*\}$.}
 \begin{theorem}\label{tight}
Suppose that problem (SDR2) is feasible. Then, there always exist optimal solution satisfying $\operatorname{rank}(\Wb^*_k) = 1, \forall k$.
\end{theorem}

{\bf Proof:} According to \cite[Theorem 3.2]{huang2010rank}, problem (SDR2) always has optimal $\{\Wb^*_k\}$ such that 
\begin{align}
   \sum_{k=1}^K \operatorname{rank}^2(\Wb^*_k) \leq K+1.
\end{align}
This implies that the rank-one optimal solution is always achieved. This completes the proof. \hfill $\blacksquare$
\\

{
The optimal $\{\Wb_k^*\}$ for (SDR2) can be attained by off-the-shelf solvers like CVX \cite{grant2014cvx}. 
After obtaining $\{\Wb_k^*\}$, according to Theorem 1, an optimal beamformer $\{\wb_k^*\}$ for (P2) can always be extracted from it, e.g., via the rank reduction technique in \cite{huang2010rank}.}

Solving problem (P2) using the SDR method provides valuable insights into its structure, revealing hidden convexity that allows us to obtain globally optimal solutions {\cite{liu2024survey}}. However, the SDR method typically entails a relatively high computational complexity \cite{luo2010semidefinite}. To facilitate practical applications, this motivates us to leverage the hidden convexity of (P2) to develop a low-complexity algorithm.



\section{Optimal Solution via Primal-Dual Method}\label{section_PD}

\subsection{Reformulation of Problem (P2)}
{
According to the optimal solution of SDR problem provided in Proposition 2 and the tightness of SDR established in Theorem 1, the optimal $\{q_n^{dl,*}\}$ for (P2) can be attained by 
\begin{align}\label{qdl}
    q_n^{dl,*} = \alpha \sum_{k=1}^K\eb_n^\top\wb_k^*(\wb_k^*)^\Hf \eb_n, \forall n \in \mathcal{N}.
\end{align}
}

By substituting \eqref{qdl} into (P2), the original J-FCBD problem boils down to the following beamforming design problem:
\begin{subequations} \label{P5}
    \begin{align}
     \text{(P3):}\  & \min _{\{\wb_k\}} \  \sum_{k=1}^K \wb_k^\Hf\wb_k \label{P5-1}\\
    \text { s.t. } 
        & \widetilde{\Gamma}_k \wb_{k}^\Hf \Hb_{k}\wb_{k} \geq  \sum_{i =1}^K \wb_{i}^\Hf \Ab_k \wb_{i} + \sigma^2_v, \forall k \in \mathcal{K}, \label{P5-2}\\
        & \sum_{k =1}^K \wb_{k}^\Hf \Bb \wb_{k} \geq \widetilde{\Gamma}_s, \label{P5-4}
    \end{align}
\end{subequations}
where $\{\Ab_k\}$ and $\Bb$ are defined under problem (SDR2). In fact, problem (SDR2) is the SDR version of (P3), omitting the positive constant $1+\alpha$ in \eqref{P5-1}, as it doesn't affect the solution.

Problem (P3) is a nonconvex quadratic constrained quadratic programming (QCQP) and challenging to solve. 
Intriguingly, we find that the main difficulty in solving (P3) comes from the nonconvex sensing SINR constraint \eqref{P5-4}.
In particular, by discarding constraint \eqref{P5-4}, the remaining problem is given by
\begin{subequations} \label{P5DL}
    \begin{align}
       &\text{(P4):}\ \min _{\{\wb_{k}\}} \  \sum_{k=1}^K \wb_{k}^\Hf\wb_{k} \\
        &\hspace{-2mm}\text { s.t. } 
         \widetilde{\Gamma}_k \wb_{k}^{\Hf} \Hb_{k}\wb_{k} \geq  \sum_{i =1}^K \wb_{i}^{\Hf} \Ab_k \wb_{i} + \sigma^2_v, \forall k \in \mathcal{K},\label{P5DL-2}
    \end{align}
\end{subequations}
which can actually be equivalently recast into a convex second-order cone programming (SOCP), and thus can be efficiently solved using the UL-DL duality with the fixed-point method \cite{liu2021uplink, fan2023qos}. 

Notably, the optimal solution, denoted by $\{\hat{\wb}_k^{*}\}$, for (P4) may already satisfy constraint \eqref{P5-4} in some cases. In these cases, $\{\hat{\wb}_k^{*}\}$ will also be optimal for (P3), indicating that the globally optimal solution of (P3) can be efficiently attained by solving a simple convex SOCP. However, in more general cases, one still needs to address (P3). 
\begin{remark}
The case where the optimal solution for (P4) also satisfies \eqref{P5-4} typically occurs in scenarios prioritizing high communication quality. 
This alignment occurs when sensing sufficiently benefits from high communication beam gain, such as with a small target-user distance or less stringent sensing requirements.
\end{remark}

{

A few existing works \cite{xu2014multiuser, attiah2024beamforming} have studied non-convex QCQP problems for beamforming design similar to (P3), which are solved using PD-based algorithms with the fixed-point method. However, the considered fronthaul compression introduces an additional complexity to our problem (P3): the interference for each user $k$ arises not only from other users’ signals but also from the design of its own signal, as indicated by the term $\wb_k^H\Ab_k\wb_k$. This added complexity impacts the application of PD-based fixed-point algorithms. While the hidden convexity discussed in Section \ref{section_SDR} allows for the application of PD-based algorithms by ensuring a zero duality gap between (P3) and its dual problem, it faces additional challenges, such as the need for more careful considerations of the conditions under which the fixed-point method ensures that the constraint with respect to dual variables hold as equality.

}

\subsection{Proposed PD-based Algorithm for (P3)}

To illustrate the PD-based algorithm, we consider the following partial dual problem of (P3): 
\begin{subequations} \label{P6}
    \begin{align}
    &\text{(D1):}\    \max_{\lambda \geq 0}  \min _{\{\wb_k\}} \ \sum_{k=1}^K \wb_k^\Hf\wb_k + \lambda\Big(\widetilde{\Gamma}_s -\sum_{k =1}^K \wb_{k}^\Hf \Bb\wb_{k} \Big)  \\
   & \text { s.t. } 
         \widetilde{\Gamma}_k \wb_{k}^\Hf \Hb_{k}\wb_{k} \geq  \sum_{i =1}^K \wb_{i}^\Hf \Ab_k \wb_{i} + \sigma^2_v, \forall k \in \mathcal{K}, \label{P6-2}
    \end{align}
\end{subequations}
where $\lambda \geq 0$ is the dual variable associated with the sensing SINR constraint \eqref{P5-4}. 

Problem (D1) is a bilevel problem that can be tackled by alternately updating the outer variable $\lambda$ and the inner variables ${\wb_k}$. This process involves solving two subproblems iteratively. For the inner problem, given the dual variable $\lambda \geq 0$, we solve the beamforming design problem, which is given by
\begin{subequations} \label{P6.1}
    \begin{align}
    \text{(D1.1):}\  \mathcal{L}_{min}(\lambda) \triangleq  \min _{\{\wb_k\}} \ & \sum_{k=1}^K \wb_k^\Hf (\Ib - \lambda\Bb)\wb_k   \\
    \text { s.t. } 
       \ & \eqref{P6-2}.
    \end{align}
\end{subequations}
{Denote $\{\wb^*_{k}(\lambda)\}$ as the optimal solution for (D1.1). For the outer problem, we update the dual variable based on $\{\wb^*_{k}(\lambda)\}$ by solving }
\begin{align} \label{P6.2}
    \text{(D1.2):}\     \max_{\lambda \geq 0} \ \lambda\left(\widetilde{\Gamma}_s  -
  \sum_{k =1}^K \wb^*_{k}(\lambda)^\Hf \Bb \wb^*_{k}(\lambda) \right). 
\end{align}

 



In what follows, we will solve problem (D1.1) and problem (D1.2), respectively. Before solving problem (D1.1), we note that the value of $\mathcal{L}_{min}(\lambda)$ is affected by its coefficient matrix $\Ib - \lambda\Bb$, which can be categorized into the following three cases:
\begin{itemize}
    \item[{\bf Case $\bf 1$}: $\Ib - \lambda\Bb \succeq \mathbf{0}$.] In this case, the objective function of (D1.1) is convex with $\{\wb_k\}$ and $\mathcal{L}_{min}(\lambda)$ must be nonnegative. In this condition, problem (D1.1) can be solved by the same method as problem (P4). 
    Since $\Ib - \lambda\Bb$ is positive semidefinite, the range of $\lambda$ must ensure that all its eigenvalues are non-negative. This range can be calculated as
    $\lambda \in \mathcal{D}_1 \triangleq \left[0, \frac{N_t}{(\alpha + N)\max_i |g_i|^2} \right]$.
    
    \item[{\bf Case $\bf 2$}: $\Ib - \lambda\Bb \preceq \mathbf{0}$.] In this case, we have $\mathcal{L}_{min}(\lambda) = - \infty$ because one can always decrease the objective value by increasing the transmit power $\{p_k \triangleq \|\wb_k\|^2 \}$ while ensuring \eqref{P6-2}. Similarly, to ensure non-positive eigenvalues, the range of $\lambda$ in this case can be calculated as $\lambda \in \mathcal{D}_2 \triangleq \left[\frac{N_t}{\alpha\min_i |g_i|^2}, +\infty\right)$.  
    
    \item[{\bf Case $\bf 3$}: $\Ib - \lambda\Bb \nsucceq \mathbf{0}$ and $\Ib - \lambda\Bb \npreceq \mathbf{0}$.] In this case, $\mathcal{L}_{min}(\lambda)$ depends on $\lambda$, and it is possible that $\mathcal{L}_{min}(\lambda)\geq 0$. However, in this case, solving problem (D1.1) is more challenging than Case $1$ owing to the non-convexity of the objective function. Similarly, the range of $\lambda$ in this case can be obtained as 
    $\lambda \in \mathcal{D}_3 \triangleq \left(\frac{N_t}{(\alpha + N)\max_i |g_i|^2},  \frac{N_t}{\alpha\min_i |g_i|^2} \right)$.  
\end{itemize}

Therefore, to find the optimal solution of problem (D1), it suffices to solve  (D1.1) and (D1.2) for $\lambda \in \mathcal{D}_1 \cup \mathcal{D}_3$. Since for the case of  {$\Ib - \lambda\Bb \succeq 0$}, problem (D1.1) can be solved by the same method as (P4), {we are committed to study} problem (D1.1) in the more challenging case with $\lambda \in \mathcal{D}_3$ in what follows.

\subsubsection{Solve Problem (D1.1) {with} $\lambda \in \mathcal{D}_3$:}
\indent 


To begin with, we first show that (D1.1) has an equivalent duality problem by proving the following proposition.
\begin{proposition} \label{pro0}
    Suppose that problem (D1.1) is feasible. Then, (D1.1) achieves the same optimal objective value as the following dual problem when the dual problem is feasible,
    \begin{subequations} \label{P7}
        \begin{align}
     {\text{\rm(D2)}}: \ &  \max _{\{\mu_k \geq 0\}} \ \sum_{k=1}^K \mu_k\sigma_v^2  \label{P7-0}\\
        \text { s.t. } 
            & \mu_k \leq \min_{\{\|\widetilde{\wb}_k\| = 1\}} \ \frac{\widetilde{\wb}_k^\Hf \Cb(\lambda, \boldsymbol{\mu})  \widetilde{\wb}_k}{\widetilde{\Gamma}_k \widetilde{\wb}_k^\Hf\Hb_{k}\widetilde{\wb}_k}, \forall k \in \mathcal{K}, \label{P7-1}
        \end{align}
    \end{subequations}
    where $\boldsymbol{\mu} \triangleq [\mu_1, \dots, \mu_K]^\top$ and $\Cb(\lambda, \boldsymbol{\mu})  \triangleq \Ib - \lambda\Bb + \sum_{i=1}^K \mu_i\Ab_i$. Moreover, the optimal beamforming direction of (D1.1), i.e., $\{\widetilde{\wb}_k^*\}$ , is obtained via solving (D2). 
\end{proposition}

{\bf Proof:} The proof is similar to that of \cite[Proposition 4]{chang2018Qos} and we omit it. {
Although the objective function of (D1.1) is non-convex, its hidden convexity can be established similarly to the findings in Section \ref{section_SDR}, ensuring that (D1.1) and its dual problem (D2) are free of duality gap.} \hfill $\blacksquare$

Based on Proposition \ref{pro0}, we can determine the beamforming direction by solving (D2) when it is feasible. In what follows, we develop a fixed-point method to find the beamforming direction and derive the optimal transmit power $p^*_k(\lambda)\triangleq \|\wb_k^*(\lambda)\|^2$ as a closed-form expression.

\subsubsubsection{1.1) Solve the Beamforming Direction Problem}
It is important to note that the dual problem (D2) is convex with respect to $\boldsymbol{\mu}$. 
This can be established by observing that the objective function \eqref{P7-0} is linear, and the constraint \eqref{P7-1} forms a convex set because its right-hand side function is the minimum of linear functions with respect to $\boldsymbol{\mu}$, which is concave. 
Therefore, any stationary point of problem (D2) is globally optimal. In the following, we develop a computationally efficient fixed-point-based algorithm to obtain a stationary point of problem (D2). 
Before that, we show that the optimal ${\widetilde{\wb}_k^*(\lambda)}$ has a simple closed-form expression with respect to $\boldsymbol{\mu}$ in the following proposition.

\begin{proposition}\label{woptimal}
    Suppose that the dual problem (D2) is feasible. Then, the optimal $\widetilde{\wb}_k^*$ is given by
    \begin{align}\label{MVDR}
            \widetilde{\wb}^*_k=  \frac{\Cb^{-1}(\lambda, \boldsymbol{\mu}) \hb_k}{\| \Cb^{-1}(\lambda, \boldsymbol{\mu}) \hb_k\|}, \forall k \in \mathcal{K},
    \end{align}
    and the optimal $\boldsymbol{\mu}^*$ satisfies 
    \begin{align}
    \Cb(\lambda, \boldsymbol{\mu}^*) \succ 0, \ \text{with}\ \lambda \in \mathcal{D}_3. \label{c2}
    \end{align}
\end{proposition}

 {\bf Proof:} The proof is relegated to Appendix \ref{D}. \hfill $\blacksquare$

By substituting $\{\widetilde{\wb}_k^*\}$ into problem (D2), one can rewrite it as
\begin{subequations} \label{P8}
    \begin{align}
\text{(D3)}:   & \max _{\{\mu_k \geq 0\}} \  \sum_{k=1}^K \mu_k\sigma_v^2  \\
    \text { s.t. } 
        & f_k(\lambda, \boldsymbol{\mu}) \triangleq \mu_k\hb_k^\Hf\Cb^{-1}(\lambda, \boldsymbol{\mu}) \hb_k \leq  \frac{1}{\widetilde{\Gamma}_k}, \forall k \in \mathcal{K}. \label{P8-1}
    \end{align}
\end{subequations}
{
Owing to the impact of compression, the dual variable $\mu_k$ is embedded within the term $\mathbf{C}(\lambda, \boldsymbol{\mu})$, rendering the function $f_{k}(\lambda, \boldsymbol{\mu})$ non-monotonic with respect to $\mu_k$. Consequently, the notion that the constraint \eqref{P8-1} is satisfied as an equality at optimality is less intuitive compared to \cite{xu2014multiuser, attiah2024beamforming}. Nevertheless, through a rigorous proof by contradiction as shown in Proposition 5, we establish that the constraint \eqref{P8-1} does, indeed, hold as equality at optimality. This pivotal insight enables the application of an efficient fixed-point method for solving $\boldsymbol{\mu}$, like \cite{xu2014multiuser, attiah2024beamforming}.

}

\begin{proposition}\label{SINRtight}
    The optimal solution of problem (D3) satisfies $\frac{\partial f_k(\lambda, \boldsymbol{\mu})}{ \partial \mu_k} \big|_{   \mu_k = \mu_k^*} \geq 0$, ensuring that the constraint \eqref{P8-1} is met with equality at the optimum, i.e., $f_k( \lambda, \boldsymbol{\mu}^*) = \frac{1}{\widetilde{\Gamma}_k}, \forall k \in \mathcal{K}$.
\end{proposition}

{\bf Proof:} 
The proof is relegated to Appendix \ref{E}. \hfill $\blacksquare$

With Proposition \ref{SINRtight}, it then follows that the optimal $\boldsymbol{\mu}^*$ can be solved by the fixed-point method \cite{rashid1998joint}. In particular, in the $(t+1)$-th iteration of the fixed-point update, $\mu_k$ is given by
\begin{align}\label{mufix}
    \mu_k^{(t+1)} & = \frac{1}{\widetilde{\Gamma}_k\hb_k^\Hf\Cb^{-1}(\lambda, \boldsymbol{\mu}) \hb_k }   \triangleq h_k(\lambda, \boldsymbol{\mu}^{(t)}), \forall k \in \mathcal{K}.
\end{align}

However, it should be pointed out that $h_k(\lambda, \boldsymbol{\mu})$ with $\lambda \in  \mathcal{D}_3$ is not a standard interference function {\cite{yates1995uplink}}, which invalidates the previous convergence proof of fixed-point iteration. Nevertheless, in the following proposition, we show that the fixed-point iterations \eqref{mufix} still converge to the optimal point.



\begin{proposition}\label{downlink_fix}
    Suppose that problem  (D2) is feasible. Then, the fixed-point method converges to the optimal $\boldsymbol{\mu}^*$ with an initial feasible point of (D3) satisfying the condition \eqref{c2}, i.e.,
    \begin{align}\label{fea3}
         f_k\left(\lambda, \boldsymbol{\mu}^{(0)}\right) \leq \frac{1}{\widetilde{\Gamma}_k}, \forall k \in \mathcal{K}, \quad \Cb(\lambda, \boldsymbol{\mu}^{(0)}) \succ 0.
    \end{align}
\end{proposition}

{\bf Proof:} The proof is relegated to Appendix \ref{F}. \hfill $\blacksquare$

With the initial feasible point, the above fixed-point-based algorithm achieves the optimal beamforming direction by substituting $\boldsymbol{\mu}^*$ into \eqref{MVDR}.
However, it is difficult to ascertain the feasibility of (D2), and thus the initial feasible points satisfied \eqref{fea3} cannot be readily found. 
{ By exploiting the novel UL-DL duality result for (D3), we propose an enhanced fixed-point-based algorithm. This innovative use of the duality result greatly enhances the algorithm's effectiveness.}

The enhanced algorithm tackles the following UL power minimization problem, which is a reverse formulation of (D3):
\begin{subequations} \label{P9}
    \begin{align}
  \text{(D4)}:  \min _{\{\mu_k \geq 0\}} \ & \sum_{k=1}^K \mu_k \sigma_v^2 \\
    \text { s.t. } 
        & f_k(\lambda, \boldsymbol{\mu}) \geq \frac{1}{\widetilde{\Gamma}_k}, \forall k \in \mathcal{K}. \label{P9.1} 
    \end{align}
\end{subequations}

This is based on the fact that problem (D4) achieves the same optimal solution  $\boldsymbol{\mu}^*$ as problem (D3), because the optimal solutions of both (D4) and (D3) are obtained when the SINR constraints are active. This proof of the activity of SINR constraints at optimum is similar to Proposition \ref{SINRtight}. Based on this, the iterative fixed-point method \eqref{mufix} is still applicable to problem (D4), starting from a feasible point of (D4) with the \eqref{c2} constraint. 
The feasible point of (D4) can be obtained from the feasible point of (P4), i.e., $\lambda = 0$. This is because $f_k\left(\lambda>0, \boldsymbol{\mu}\right) \geq f_k\left(\lambda=0, \boldsymbol{\mu}\right), \forall k \in \mathcal{K}$ always hold. Moreover, the initial point $\boldsymbol{\mu}^{(0)}$ should be sufficient large to satisfy condition \eqref{c2} while ensuring $f_k\left(\lambda=0, \boldsymbol{\mu}^{(0)}\right) \geq \frac{1}{\widetilde{\Gamma}_k}, \forall k \in \mathcal{K}$.
The proof of convergence for the enhanced algorithm is presented in the following proposition.
 
\begin{proposition} \label{uplink_fix}
    {Suppose that problem  (D2) is feasible.} the fixed-point iteration converges to optimal $\boldsymbol{\mu}^*$ with an initial feasible point of (D4) satisfying the condition \eqref{c2}, i.e.,
    \begin{align}
             f_k\left(\lambda, \boldsymbol{\mu}^{(0)}\right) \geq \frac{1}{\widetilde{\Gamma}_k}, \forall k \in \mathcal{K}, \quad \Cb(\lambda, \boldsymbol{\mu}^{(0)}) \succ 0.
    \end{align}
\end{proposition}

{\bf Proof:} The proof is relegated to Appendix \ref{G}. \hfill $\blacksquare$

{
By applying the fixed-point method to problem (D4), we can infer from Proposition \ref{uplink_fix} that the optimal $\boldsymbol{\mu}^*$ can be determined when (D2) is feasible. The PD-based fixed-point algorithm for (D1.1) is detailed in Algorithm \ref{PDalgorithm}. During each iteration, we verify condition \eqref{c2} to ensure the solution remains feasible for (D2).
Once the optimal $\boldsymbol{\mu}^*$ is obtained by the above fixed-point iterations, the optimal beamforming direction $\{\widetilde{\wb}^*_k(\lambda)\}$ can be obtained by \eqref{MVDR}. Additionally, if (D2) is infeasible, Algorithm \ref{PDalgorithm} will fail to converge to a solution that satisfies \eqref{c2}, and the primal problem (D1.1) is unbounded from below, i.e., $\mathcal{L}_{min}(\lambda) = -\infty$ \cite{Matousek2007linear}.}

\subsubsubsection{1.2) Solve the Transmit Power Problem}
After obtaining $\{\widetilde{\wb}^*_k(\lambda)\}$, what remains for solving (D1.1) is to obtain the optimal transmission power. Define $\pb(\lambda) \triangleq [p_1(\lambda), \dots, p_K(\lambda)]^\top$, the optimal $\pb^*(\lambda)$ can be attained by the following proposition.

\begin{proposition}\label{power}
    Suppose that problem (D2) is feasible. Then, the optimal $\pb^*$ of (D1.1) is 
    \begin{align}\label{p}
        \pb^*(\lambda) = \sigma_v^2 \Sb^{-1}\mathbf{1},
    \end{align}
    where the $(k, i)$-th element of $\Sb \in \mathbb{R}^{K \times K}$ is given by 
    \begin{align}
        \Sb_{k,i} = \left\{\begin{array}{llcc}
          \widetilde{\wb}^*_{k}(\lambda)^\Hf \left(\widetilde{\Gamma}_k \Hb_{k} - \Ab_k \right)\widetilde{\wb}^*_{k}(\lambda),   & k = i, \\
          -\widetilde{\wb}^*_{i}(\lambda)^\Hf \Ab_k\widetilde{\wb}^*_{i}(\lambda),   & k \neq i.
        \end{array}\right. \notag
    \end{align}
\end{proposition}

{\bf Proof:} The proof is relegated to Appendix \ref{H}. \hfill $\blacksquare$

Based on Proposition \ref{power}, the optimal solution of (D1.1) for given $\lambda$ can be obtained as $\wb_k^*(\lambda) = \sqrt{p_k^*(\lambda)} \widetilde{\wb}_k^*(\lambda), \forall k \in \mathcal{K}$. Following this idea, we develop a PD-based algorithm for (D1.1), and its details are summarized in Algorithm \ref{PDalgorithm}.

\begin{algorithm}[t]
\caption{Proposed PD-based Algorithm for (D1.1)}
\label{PDalgorithm}
\begin{algorithmic}[1]
\REQUIRE $M, K, C^{dl}, \Gamma_k, C^{ul}, \Gamma_s, B, \hb_k, g, \sigma_v^2, \sigma_z^2, \varepsilon$
\STATE Initialize $\lambda$, $t = 0$ and $\{\mu_k^{(0)}\}$.
\REPEAT
    \IF{Condition \eqref{c2} fails}
        \STATE Exit the algorithm.
    \ENDIF 
    \STATE $t = t+1$
    \STATE Update $\{\mu_k^{(t)}\}$ by fixed-point method \eqref{mufix}.
\UNTIL{$|\mu_k^{(t)} - \mu_k^{(t-1)}| \leq \varepsilon, \forall k \in \mathcal{K}$.}
\STATE Set $\mu^*_k = \mu_k^{(t)}, \forall k \in \mathcal{K}$.
\STATE Obtain $\widetilde{\wb}_k^*(\lambda)$ by \eqref{MVDR} and $\pb^*(\lambda)$ by \eqref{p}.
\STATE Compute the beamforming as $\wb_k^*(\lambda) = \sqrt{p_i^*(\lambda)}\widetilde{\wb}_k^*(\lambda)$.
\ENSURE $\{\wb_k^*(\lambda)\}$. 
\end{algorithmic}
\end{algorithm}

\vspace{3mm}
\subsubsection{Solve Problem (D1.2)}
To find the optimal dual variable $\lambda^*$, we utilize bisection to solve problem (D1.2) according to the obtained $\{\wb_k^*(\lambda)\}$ by solving problem (D1.1). Specifically, the dual variable $\lambda$ can be updated according to its subgradient of objective function \eqref{P6.1}, which is given by $\Delta(\lambda) \triangleq \widetilde{\Gamma}_s -\sum_{k =1}^K \wb^*_{k}(\lambda)^\Hf \Bb\wb^*_{k}(\lambda)$.

Moreover, if problem (D2) is infeasible, $\{\wb_k^*(\lambda)\}$ cannot be obtained by Algorithm \ref{PDalgorithm} and problem (D1.1) is unbounded from below, i.e., $\mathcal{L}_{min}(\lambda) = -\infty$. In this case, it can be shown that $\lambda^* \leq \lambda$. This can be proved by contradiction. Given $\widetilde{\lambda} \geq \lambda$, it is not difficult to show that problem (D1.1) with $\widetilde{\lambda}$ has the same feasible set as that with $\lambda$, and a smaller objective value than that with $\lambda$, i.e., $\mathcal{L}_{min}(\widetilde{\lambda}) \leq \mathcal{L}_{min}(\lambda) = - \infty$.
Therefore, the dual variable greater than $\lambda$ cannot be optimal.

\subsection{Proposed PD-based Algorithm for J-FCBD (P1)}
Now, we present the algorithm for solving the original J-FCBD problem (P1). The optimal fronthaul compression variables are given in \eqref{qul} and \eqref{qdl}. The optimal beamforming can be found by solving (D1) via PD-based algorithm. The proposed algorithm first fixes the dual variable $\lambda$ and finds the corresponding optimal beamforming $\{\wb_k^*(\lambda)\}$ via fixed-point methods, as shown in Algorithm \ref{PDalgorithm}. According to $\{\wb_k^*(\lambda)\}$, the subgradient of $\lambda$ can be calculated, and then the dual variable $\lambda$ can be updated by bisection. Following this idea, the proposed PD-based algorithm is summarized in Algorithm \ref{sumalgorithm}.

\begin{algorithm}[t]
\caption{Proposed PD-based Algorithm for (P1)}
\label{sumalgorithm}
\begin{algorithmic}[1]
\REQUIRE $M, K, C^{dl}, \Gamma_k, C^{ul}, \Gamma_s, B, \hb_k, g, \sigma_v^2, \sigma_z^2, \varepsilon$
\STATE Obtain $\{\hat{\wb}_k^*\}$ by solving (P4).
\IF {$\{\hat{\wb}_k^*\}$ satisfies \eqref{P5-4}}
    \STATE $\wb_k^* = \hat{\wb}_k^*, \forall k.$
\ELSE 
    \STATE Initialize $\lambda_{min}$ and $\lambda_{max}$.
    \REPEAT 
        \STATE $\lambda = \frac{\lambda_{min} + \lambda_{max}}{2}$.
        \STATE Obtain $\wb_k^*(\lambda)$ by Algorithm \ref{PDalgorithm} and compute $\Delta(\lambda)$.     
        \IF{Algorithm \ref{PDalgorithm} fails or $\Delta(\lambda) < 0$}
            \STATE Update $\lambda_{max} = \lambda$.
        \ELSE
        \STATE Update $\lambda_{min} = \lambda$.
        \ENDIF
    \UNTIL{$|\Delta(\lambda)| \leq \varepsilon$.}
    \STATE Set $\lambda^* = \lambda$ and $\wb_k^* = \wb_k^*(\lambda^*), \forall k \in \mathcal{K}$.
\ENDIF
\STATE Obtain the UL compression noise levels $q_{m}^{ul, *}$ by \eqref{qul} and the DL compression noise levels $q_{m}^{dl, *}$ by \eqref{qdl}.
\ENSURE $\{\wb_k^*, q_{m}^{ul, *}, q_{m}^{dl, *}\}$. 
\end{algorithmic}
\end{algorithm}

\section{Simulation Results}\label{section_simulation}
In this section, extensive simulations are conducted to validate the performance of our proposed algorithms. Unless specified otherwise, we consider a networked ISAC system with $L = 2$ TXs and one RX. Each TX and RX is equipped with $N_t = M = 32$ antennas. The number of communication users is set to $K = 16$. The carrier frequency, channel bandwidth, and noise power spectrum density are set to $f_c = 3 \operatorname{GHz}$, $B = 10 \operatorname{MHz}$, and -174 $\operatorname{dBm/Hz}$, respectively. The path loss of the communication/sensing channel is given by $128.1 + 37.6\log_{10}(d)$, where $d$ represents the signal propagation distance \cite{lte2011evolved}. The users are randomly located within a range of 500 m from the TXs. The target is randomly located within a range of 500 m from both TXs and RX. The fronthaul capacity is set to $C^{dl} = C^{ul} = 30$ Mbps. The SINR requirements for sensing and communication are set to $\Gamma_s = 10 \operatorname{dB}$ and $\Gamma_c \triangleq \Gamma_k = 10 \operatorname{dB}, \forall k \in \mathcal{K}$, respectively.

\subsection{Convergence of the Proposed PD-Based Algorithm}
\begin{figure}[!t]   
\centering{
    \includegraphics[width=0.96\linewidth]{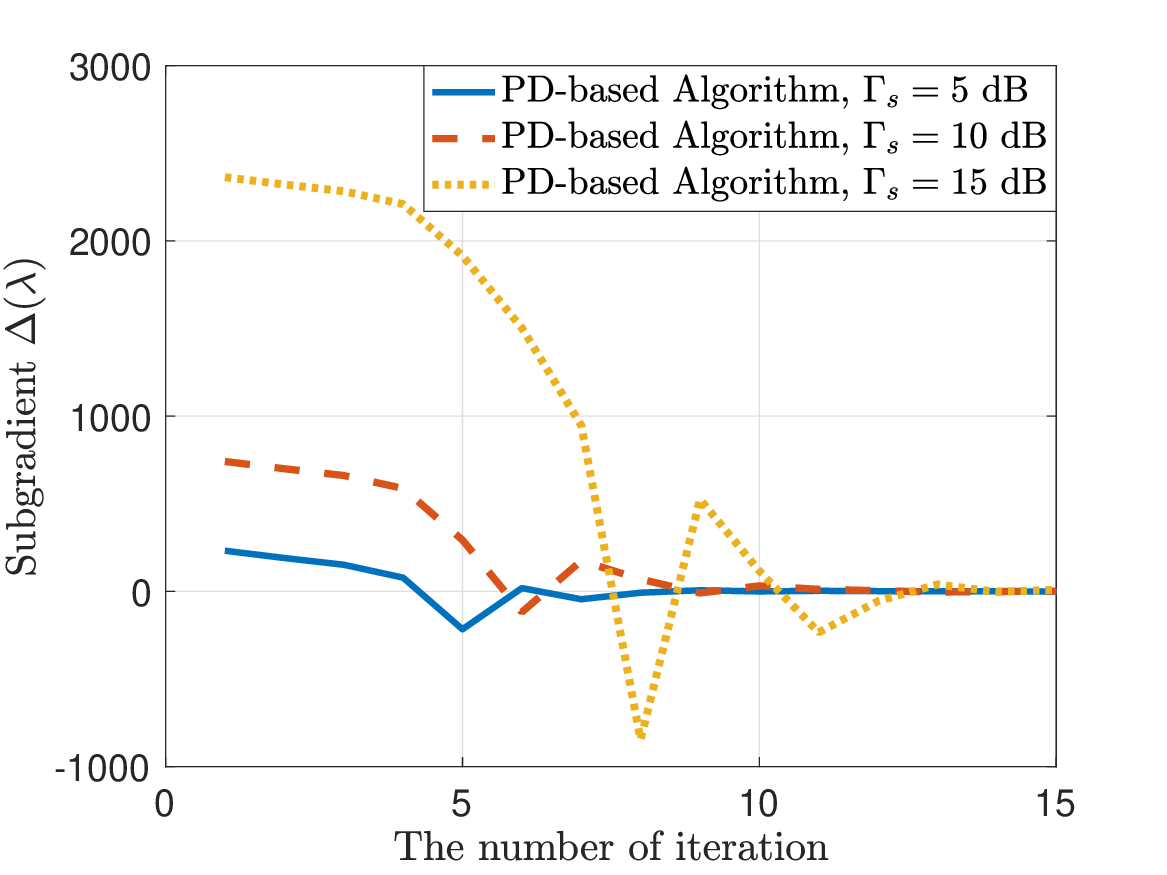}}
    \caption{\small The convergence performance of the proposed PD-based algorithm.}
    \label{fig_convergence}
\end{figure}

 In Fig. \ref{fig_convergence}, we evaluate the convergence behaviors of the proposed PD-based algorithm. The convergence of the proposed algorithm occurs after fewer iterations, e.g., with $\Gamma_s = \Gamma_c = 10 \operatorname{dB}$, the number of iterations to reach convergence is $t = 10$. An interesting phenomenon in Fig.\ref{fig_convergence} is that the proposed algorithm converges a little slower with more stringent sensing SINR requirement $\Gamma_s$. 
This is because when $\Gamma_s$ increases,  the magnitude of the subgradient $|\Delta(\lambda)|$ for the same $\lambda$ will increase, resulting in more iterations required to reach the convergence accuracy $|\Delta(\lambda)| \leq \varepsilon = 0.01$. 

\begin{figure}[!t]   
\centering
    \includegraphics[width=0.96\linewidth]{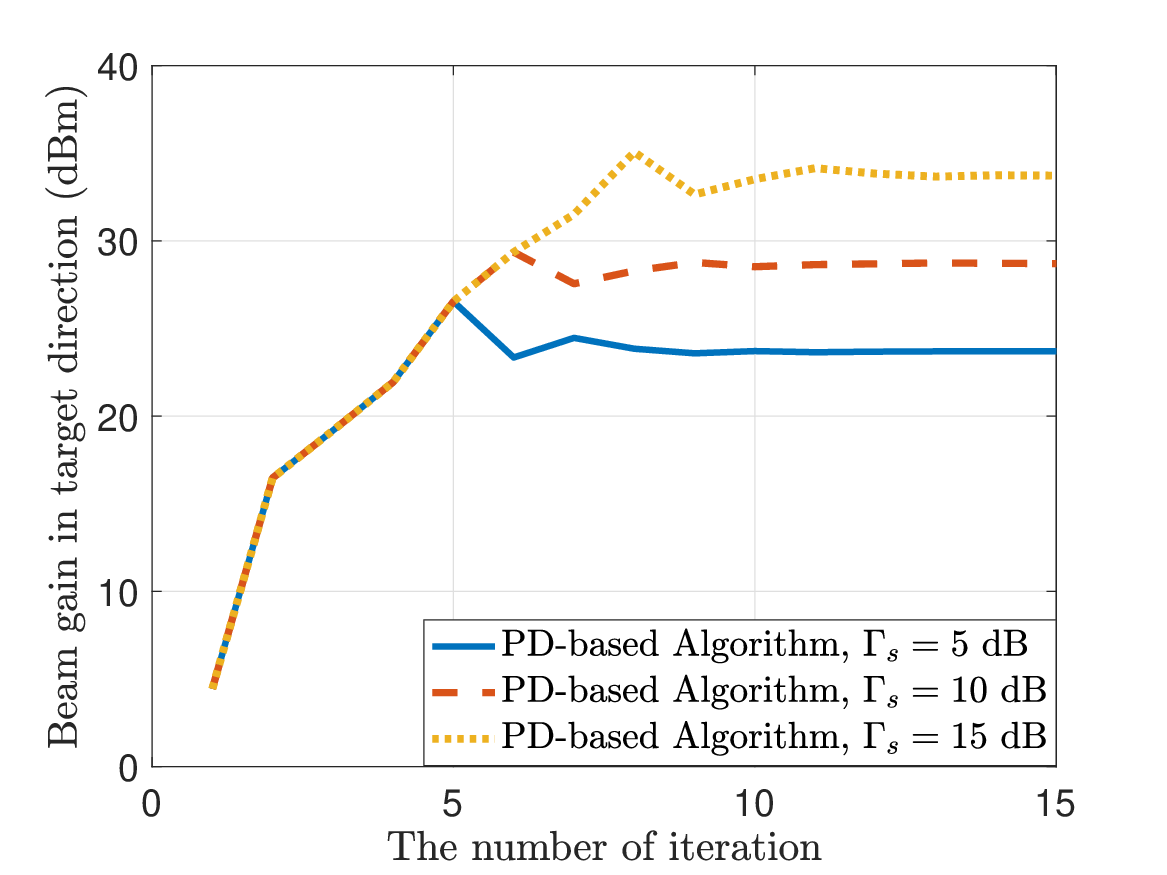}
    \caption{\small The beam gain in target direction versus the number of iterations with different sensing SINRs. }
    \label{fig_power_target}
\end{figure}

In Fig. \ref{fig_power_target}, the beam gain in target direction versus iterations is illustrated. One can observe that in the iteration, the beam gain in the target direction increases first and then oscillates in a small range until convergence. 
This is because, to meet sensing SINR requirement, the dual variable $\lambda$ needs to first increase from $\frac{\lambda_{min} + \lambda_{max}}{2}$ via bisection to obtain more beam gain in target direction. When the sensing requirement is satisfied, i.e., $\Delta(\lambda) \leq 0$, the $\lambda$ is searched in a small range to reach the convergence accuracy. Furthermore, more gain will be allocated to the target direction to meet a more stringent sensing SINR requirement.

\subsection{Effectiveness and Efficiency of the Proposed PD-Based Algorithm}
In this subsection, to illustrate the effectiveness and efficiency of the proposed PD-based algorithm, we compare it with the SDR-based algorithm and a suboptimal algorithm. The suboptimal algorithm designs communication and sensing separately. Specifically, it first solves (P4) to obtain optimal $\{\hat{\wb}_{k}^*\}$ via UL-DL duality \cite{liu2021uplink}. Subsequently, to satisfy the sensing requirement \eqref{P5-4}, the solution $\{\hat{\wb}_{k}^*\}$ is scaled by a factor $\gamma = \operatorname{max}\Big\{1, \frac{\widetilde{\Gamma}_s}{\sum_{k=1}^K (\hat{\wb}_{k}^*)^\Hf \Bb \hat{\wb}_{k}^*}\Big\}$.

\begin{figure}[!t]   \centering
    \includegraphics[width=0.96\linewidth]{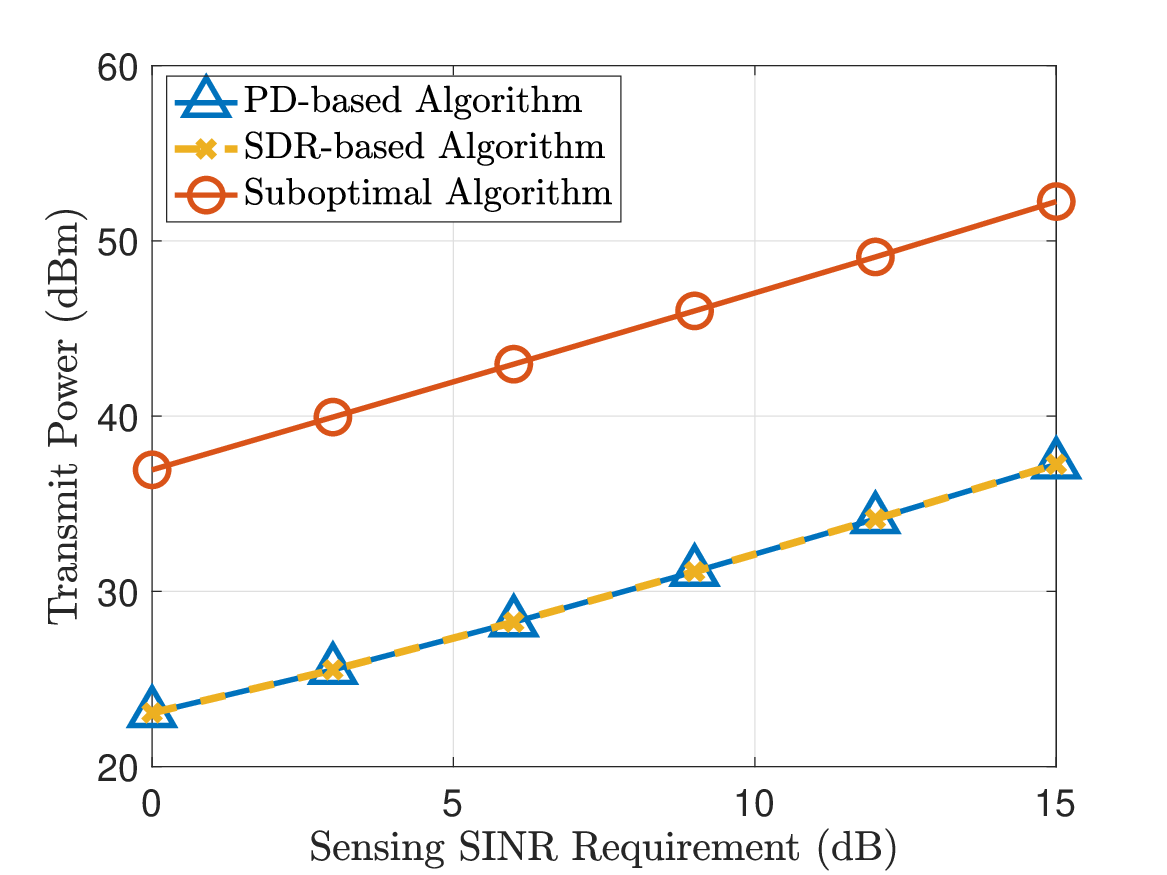}
    \caption{\small The power consumption versus sensing SINR requirement.}
    \label{fig_compare_solution}
\end{figure}

To show the effectiveness of proposed algorithm, Figs. \ref{fig_compare_solution} plots the power consumption of three algorithms. We can see from Figs.\ref{fig_compare_solution} that the proposed PD-based algorithm returns the same solution with SDR-based algorithm and significantly outperforms the suboptimal algorithm. This verifies the global optimality of the solution returned by the proposed PD-based algorithm. Moreover, intuitively, as sensing SINR requirement $\Gamma_s$ increases, power consumption also increases.

\begin{figure}[!t]   \centering
    \includegraphics[width=0.96\linewidth]{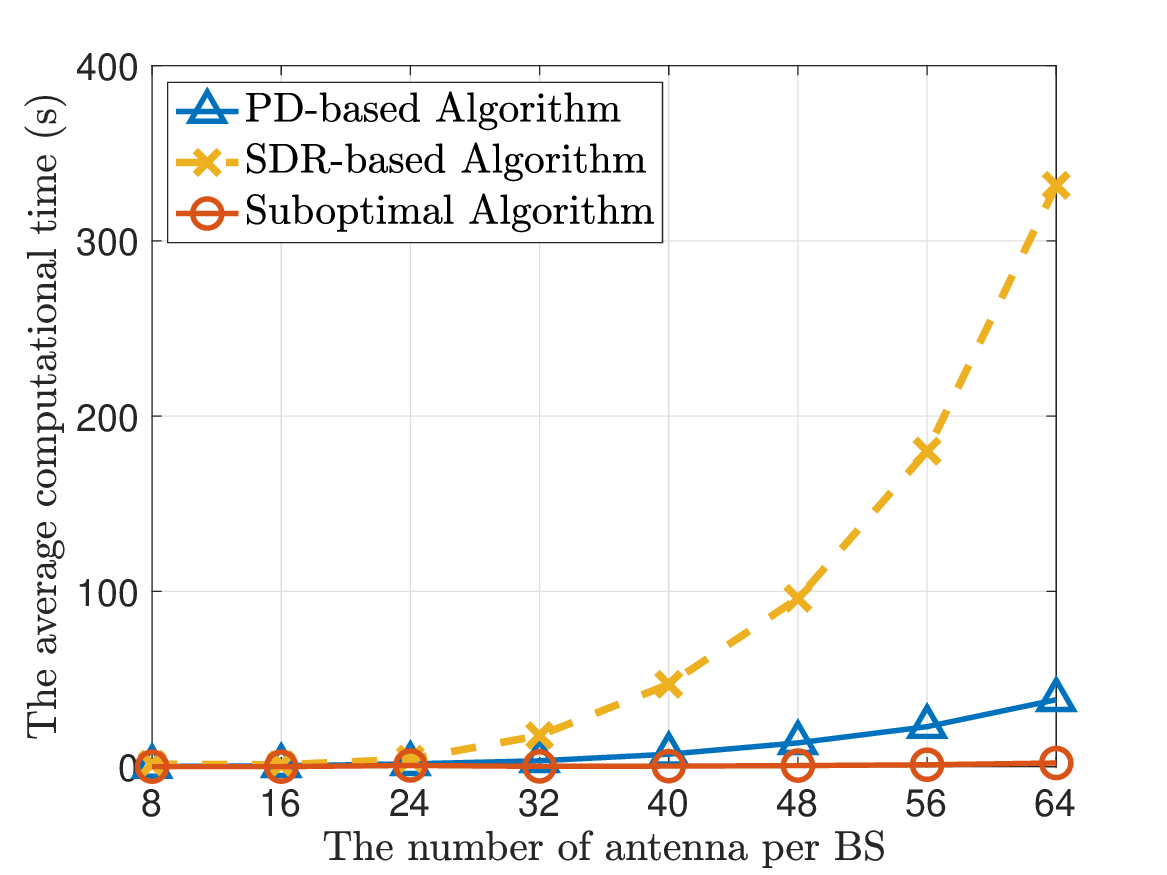}
    \caption{\small The average computation time (in second) comparison of different algorithms versus the number of antennas per BS.}
    \label{fig_compare_time}
\end{figure}

To show the efficiency of proposed algorithm, Fig. \ref{fig_compare_time} plots the average computational times of different algorithms versus the number of antenna per BS (i.e., $N_t, M$) with $K = \frac{N_t}{2}$. We conduct the simulations by using MATLAB 2023b on a computer with Intel Core i7-13700K CPU and 64GB RAM. Compared with SDR-based algorithm, PD-based algorithm achieves greater time reduction of the averaged computational time, especially when the number of antennas is large. Specifically, with $N_t \geq 40$, PD-based algorithm 
 reduces the computational time by more than 80 $\%$ compared with SDR-based algorithm.  It also can be observed that the running time of the proposed algorithm is larger than the suboptimal algorithm. This is the consumption caused by performance optimization.

\subsection{Impact of Fronthaul Capacity}
\begin{figure}[!t]   \centering
    \includegraphics[width=0.96\linewidth]{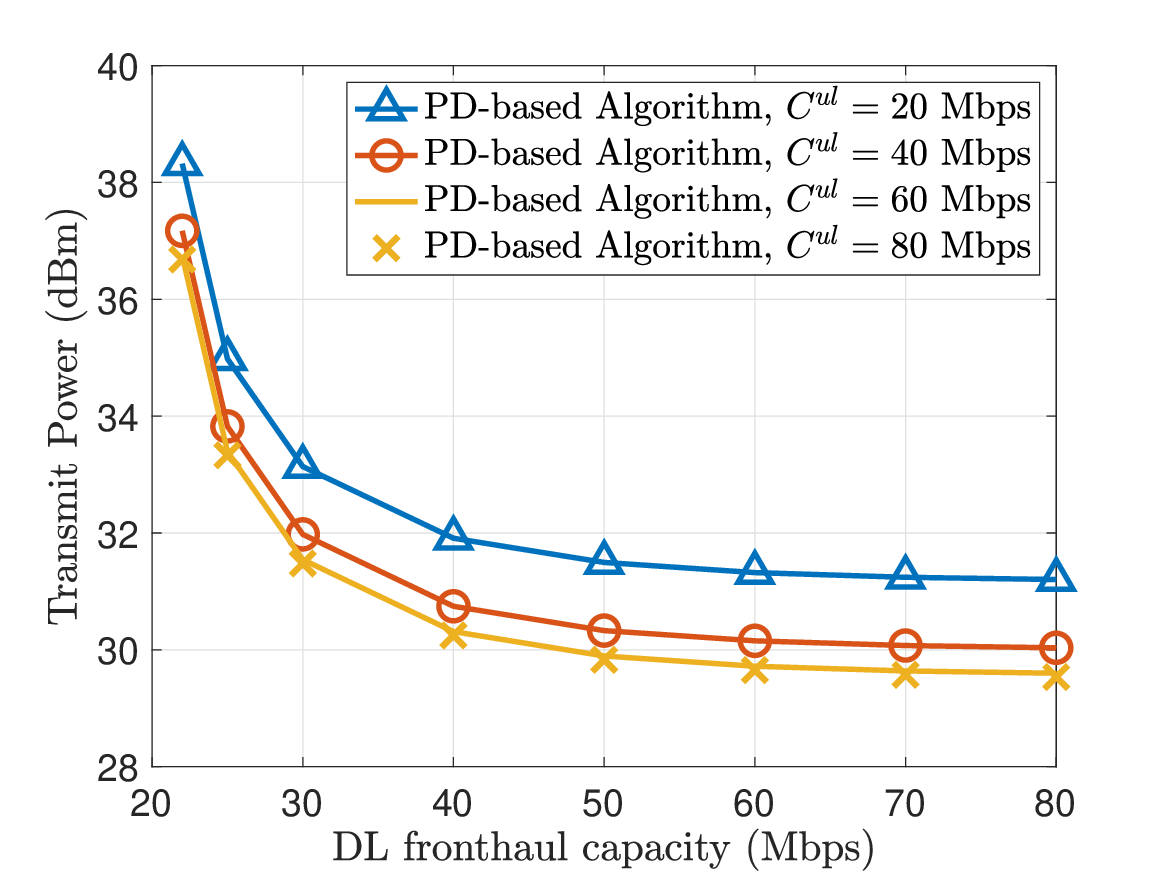}
    \caption{\small The power consumption versus DL fronthaul capacity with different UL fronthaul capacity.}
    \label{fig_fronthaul}
\end{figure}

In Fig. \ref{fig_fronthaul}, the impact of DL/UL fronthaul capacity on power consumption is illustrated.
As $C^{dl}$ increases, the transmit power consumption will decrease. This is because with larger $C^{dl}$, the DL compression noise interferes less with communication and thus the communication SINR requirement will be easier to meet. 
The optimal beam will be directed more in the target direction, completing communication and sensing with lower power. Similarly, as $C^{ul}$ increases, the transmit power consumption will also decrease because the UL compression noise interferes less with the sensing echo signal. Moreover, when the fronthaul capacity is greater than a certain value, i.e., $C^{dl} = C^{ul} = 60 \operatorname{Mbps}$, further increases in the fronthaul capacity only brings a marginal improvement to the power saving, as the capacity is no longer a bottleneck.

\section{Conclusion}\label{section_conclusion}
In this paper, we {have} studied the J-FCBD in a fronthaul-capacity-limited networked ISAC system. The J-FCBD optimization {aims} to minimize the total transmit power while satisfying the SINR requirements for both communication and sensing, as well as the fronthaul capacity constraints. We first derived the optimal fronthaul compression variables in closed forms along with the beamformers. Subsequently, we proposed an SDR-based algorithm to solve the problem and demonstrated its global optimality by proving the tightness of the SDR problem. To address the high complexity of the SDR-based algorithm, we further developed a computationally efficient PD-based algorithm using the duality results from the tightness of SDR, which also {achieves} global optimality. Simulation results showed that, compared to the SDR-based algorithm, the proposed PD-based algorithm maintained the same performance in power saving and significantly reduced computational time, especially when the number of antennas was large. Additionally, simulation results showed that the impact of fronthaul capacity on performance could not be neglected, proving the necessity of considering the limited fronthaul capacity.


%

\appendices
\section{Proof of Proposition \ref{lemmaul}} \label{A}
    In problem (P1), the variable ${q}_m^{ul}$ affects only the UL fronthaul constraint \eqref{P1-3} and the sensing SINR constraint \eqref{P1-4}.
    These constraints can be equivalently expressed as
        \begin{align}
            {q}_m^{ul} & \geq  \frac{\beta}{M}  \ab_{t}^\Hf \boldsymbol{\Sigma}_g \Rb \boldsymbol{\Sigma}_g^\Hf \ab_{t} + \beta\sigma_z^2, \forall m \in \mathcal{M}, \label{P1-4-1}\\
            \operatorname{SINR}^s &\triangleq \ab_{t}^\Hf\boldsymbol{\Sigma}_g \Rb \boldsymbol{\Sigma}_g^\Hf\ab_{t}\sum_{m = 1}^M \frac{1}{{q}_m^{ul} + \sigma^2_z} \geq \Gamma_s. \label{P1-5-1}
        \end{align}
    From \eqref{P1-5-1}, we see that the sensing SINR, i.e., $\operatorname{SINR}^s$, is monotonically decreasing with $\{q_m^{ul}\}$. 
    Suppose that, given a solution $\{q_m^{ul}\}$, the corresponding constraints in \eqref{P1-4-1} holds with inequality. Then, one can always decrease $\{q_m^{ul}\}$ and $\Rb$ accordingly, such that constraint \eqref{P1-5-1} is still satisfied and the objective value of problem (P1) decreases, until all the constraints in \eqref{P1-4-1} hold with equalities. As a result, the optimal closed-form solutions of $\{q_m^{ul}\}$ are obtained. This completes the proof. \hfill $\blacksquare$

\section{Proof of Proposition \ref{lemmadl}} \label{B}
    Suppose that $\{q_n^{dl,*}\}$ and $\{\Wb^*_k\}$ are a pair of optimal solutions for (SDR1), which render constraints \eqref{SDR2-1} inactive while satisfying constraints \eqref{SDR2-1}-\eqref{SDR2-5}. Then, one can find another pair of solutions $\{\widetilde{q}_n^{dl}\}$ and $\{\widetilde{\Wb}_k\}$ that make \eqref{SDR2-1} active, while maintaining the same objective value and ensuring constraints \eqref{SDR2-2}-\eqref{SDR2-4} are satisfied. Let's elaborate on this as follows. 
    
    Define $\delta_n \triangleq q_{n}^{dl,*} - \alpha \sum_{k=1}^K \eb_n^\top\Wb^*_k\eb_n \geq 0, \forall n \in \mathcal{N}$, and $\boldsymbol{\Lambda}_{\delta} \triangleq \operatorname{diag}(\delta_1, \dots, \delta_N)$. Then, $\Qb_{dl}^{*}$ can be expressed as
    \begin{align}\label{Qassume}
       \Qb_{dl}^{*} = \alpha\operatorname{diag}\left(\sum_{k=1}^K \Wb_{k}^*\right) + \boldsymbol{\Lambda}_{\delta}.
    \end{align}
    With $\{\Wb_{k}^*, \Qb_{dl}^{*}\}$, let's construct another pair of solutions $\{\widetilde{\Wb}_k, \widetilde{\Qb}^{dl}\}$ that make constraint \eqref{SDR2-1} active, which can be written as
    \begin{align} 
        \widetilde{\Wb}_k &= \Wb_k^* + \alpha_k\boldsymbol{\Lambda}_{\delta}, \forall k \in \mathcal{K}, \label{newsolution1}\\
        \widetilde{\Qb}_{dl} &=\alpha\operatorname{diag}\left(\sum\limits_{k=1}^K \widetilde{\Wb}_{k}\right).\label{newsolution2}
    \end{align}
    Here, $ \alpha_k \geq 0$ with $\sum_{k=1}^K \alpha_k = \frac{1}{1+\alpha}$. From \eqref{Qassume}, \eqref{newsolution1} and \eqref{newsolution2}, we have 
    \begin{align}
        \widetilde{\Qb}_{dl} = \Qb_{dl}^{*} - \frac{1}{1+\alpha}\boldsymbol{\Lambda}_{\delta}.
    \end{align}
    Now, it is not difficult to verify that 
    \begin{align}
        \sum_{k=1}^K \Wb_{k}^* + \Qb_{dl}^* = \sum_{k=1}^K \widetilde \Wb_{k} + \widetilde \Qb_{dl},
    \end{align}
    which indicates that the solutions $\widetilde \Qb_{dl}$ and $\{\widetilde \Wb_{k}\}$ achieve the same objective value as $\Qb_{dl}^*$ and $\{\Wb_{k}^*\}$, and the constraint \eqref{SDR2-4} is also satisfied. The remaining is to check whether $\widetilde \Qb_{dl}$ and $\{\widetilde \Wb_{k}\}$ can still satisfy with constraint \eqref{SDR2-2}. To this end, we 
    substitute $\widetilde \Qb_{dl}$ and $\{\widetilde \Wb_{k}\}$ into \eqref{SDR2-2} and obtain
    \begin{align}
      \operatorname{SINR}_{k}  (\widetilde{\Wb}_k, \widetilde{\Qb}^{dl}) &= \frac{ \hb_{k}^\Hf\left(\Wb^*_{k} + \alpha_k \boldsymbol{\Lambda}_{\delta} \right) \hb_{k}^\Hf }{ \hb_{k}^\Hf \left( \Rb - \Wb^*_{k} - \alpha_k\boldsymbol{\Lambda}_{\delta} \right)\hb_{k}  + \sigma^2_v} \notag\\
      & \geq \operatorname{SINR}_{k}(\Wb^*_k, \Qb^{dl, *}),
    \end{align}
    which shows that $\widetilde \Qb_{dl}$ and $\{\widetilde \Wb_{k}\}$ achieve higher communication SINRs and thus satisfiy \eqref{SDR2-2}.

    Summarily, if problem (SDR1) is feasible, there always exist optimal solutions such that the constraints \eqref{SDR2-1} hold with equalities. This completes the proof.
    \hfill $\blacksquare$

\section{Proof of Proposition \ref{woptimal}} \label{D}
    The optimal $\wb_k^*$ of (D2) can be derived by solving 
     \begin{align}\label{pro_w_k}
       h_k(\lambda, \boldsymbol{\mu}, \widetilde{\wb}_k^*) \triangleq \min_{\{\|\widetilde{\wb}_k\| = 1\}} \ \frac{\widetilde{\wb}_k^\Hf \Cb(\lambda, \boldsymbol{\mu})  \widetilde{\wb}_k}{\widetilde{\Gamma}_k \widetilde{\wb}_k^\Hf\hb_{k}\widetilde{\wb}_k}, \forall k \in \mathcal{K}.
    \end{align} 
    From the constraints of (D2), the optimal value $h_k(\lambda, \boldsymbol{\mu}, \widetilde{\wb}_k^*)$ must satisfy the following condition,
    \begin{align}\label{fea_C}
       h_k(\lambda, \boldsymbol{\mu}, \widetilde{\wb}_k^*) \geq \mu_k \geq 0, \forall k \in \mathcal{K}.
    \end{align} 
    
    With $\lambda \in \mathcal{D}_3$, we analyze the optimal $\widetilde{\wb}_k^*$ and the feasible condition \eqref{fea_C} in three cases:
    
    Firstly, if the coefficient matrix of problem \eqref{pro_w_k} satisfies $\Cb(\lambda, \boldsymbol{\mu}^*) \not\succeq 0$, then $h_k(\lambda, \boldsymbol{\mu}^*, \widetilde{\wb}_k^*)<0, \forall k \in \mathcal{K}$. In this case, there does not exist $\mu_k$ that satisfies the condition \eqref{fea_C}.
    
    Secondly, if the matrix $\Cb(\lambda, \boldsymbol{\mu}^*) \succeq 0$ and $\Cb(\lambda, \boldsymbol{\mu}^*) \nsucc 0$, then $h_k(\lambda, \boldsymbol{\mu}^*, \widetilde{\wb}_k^*)=0, \forall k \in \mathcal{K}$. There exists only one feasible $\mu_k$ that satisfies the condition \eqref{fea_C}, which is $\mu_k = 0, \forall k \in \mathcal{K}$. However, this induces a contradiction to $\Cb(\lambda, \boldsymbol{\mu}^*) \succeq 0$. 
    
    Finally, if the matrix $\Cb(\lambda, \boldsymbol{\mu}^*) \succ 0$, then $h_k(\lambda, \boldsymbol{\mu}^*, \widetilde{\wb}_k^*) > 0, \forall k \in \mathcal{K}$. In this case, the optimal $\widetilde{\wb}_k^*$ can be obtained as the well-known MVDR beamforming,
    \begin{align}\label{MVDR1}
        \widetilde{\wb}^*_k =  \frac{\Cb^{-1}(\lambda, \boldsymbol{\mu}) \hb_k}{\| \Cb^{-1}(\lambda, \boldsymbol{\mu}) \hb_k\|}, \forall k \in \mathcal{K}.
    \end{align} 
    The proof is thus completed.
    \hfill $\blacksquare$

\section{Proof of Proposition \ref{SINRtight}}\label{E}
    We first show that $\frac{\partial f_k(\lambda, \boldsymbol{\mu}) }{ \partial \mu_k} \big|_{   \mu_k = \mu_k^*} \geq 0$ should be satisfied at the optimum of  (D3).
    The derivative of the function $f_k(\lambda, \boldsymbol{\mu})$ with respect to $\{\mu_i\}$ can be calculated as
    \begin{align}
        \frac{\partial f_k }{\partial \mu_k} 
        & = \hb_k^\Hf \Cb^{-1}\Big(\Ib - \lambda \Bb + \sum_{i \neq k}^K \mu_i\Ab_{i}   \Big)\Cb^{-1}\hb_k, \label{muk}\\
        \frac{\partial f_k }{\partial \mu_i} &= -\mu_k\Tr\left(\Cb^{-1}\hb_k \hb_k^\Hf \Cb^{-1}\Ab_i \right) \leq 0, \ \forall i \neq k,
        \label{mui}
    \end{align}
    where $f_k$ and $\Cb$ are simplified expressions for $f_k(\lambda, \boldsymbol{\mu})$ and $\Cb(\lambda, \boldsymbol{\mu})$, respectively.
    
    From \eqref{muk}, the monotonicity of $f_k(\lambda, \boldsymbol{\mu})$ for $\mu_k$ is challenging to determine directly.

    In the following, we prove that $\frac{\partial f_k(\lambda, \boldsymbol{\mu}) }{ \partial \mu_k} \big|_{   \mu_k = \mu_k^*} \geq 0$ by contradiction.
    Suppose that there exists at least one $k$ such that $\frac{\partial f_k }{\partial \mu_k}\big|_{  \mu_k = \mu_k^*} \leq 0$. Define this set as $\bar{\mathcal{K}}$.
    Given a small positive constant as $\delta$, we construct a new solution for (D3) as 
    \begin{align}
        \widetilde{\mu}_k = \mu_k^* + \delta, \forall k \in \bar{\mathcal{K}}, \quad
            \widetilde{\mu}_i = \mu_i^*, \forall k \in \mathcal{K} /\bar{\mathcal{K}},
    \end{align}
    which achieves larger objective value than $\boldsymbol{\mu}^*$. Moreover, the solution $\widetilde{\boldsymbol{\mu}}$ is feasible for (D3) since $\frac{\partial f_k }{\partial \mu_k}\big|_{  \mu_k = \mu_k^*} \leq 0, \forall k \in \bar{\mathcal{K}}$ and \eqref{mui} lead to $f_k(\lambda, \widetilde{\boldsymbol{\mu}}) \leq f_k(\lambda, \boldsymbol{\mu}^*) \leq \frac{1}{\widetilde{\Gamma}_k}, \forall k \in \mathcal{K}$.
    This contradicts the assumption that the solution $\boldsymbol{\mu}^*$ with $\frac{\partial f_k }{\partial \mu_k}\big|_{\mu_k = \mu_k^*} \leq 0, \forall k \in \bar{\mathcal{K}}$ is optimal for (D3). 
    Therefore, the optimal $\boldsymbol{\mu}^*$ should satisfy 
    \begin{align}\label{partial}
        \frac{\partial f_k}{ \partial \mu_k} \Big|_{   \mu_k = \mu_k^*} \geq 0, \forall k \in \mathcal{K}.
    \end{align}
    Based on this conclusion, the optimal $\boldsymbol{\mu}^*$ should be obtained when all constraints \eqref{P8-1} are met with equality. 
    This is because failure to satisfy these conditions would enable further improvement in the objective value of (D3) by increasing $\boldsymbol{\mu}$. The proof is thus completed.
    \hfill $\blacksquare$

\section{Proof of Proposition \ref{downlink_fix}}\label{F}
    Under the consumption that (D2) is feasible, we can initialize the fixed point iteration \eqref{mufix} with a feasible point $\boldsymbol{\mu}^{(0)}$ of (D3) satisfying the condition \eqref{c2}, i.e.,
    \begin{align}\label{fea1}
         f_k\left(\lambda, \boldsymbol{\mu}^{(0)}\right) \leq \frac{1}{\widetilde{\Gamma}_k}, \forall k \in \mathcal{K}, \quad \Cb(\lambda, \boldsymbol{\mu}^{(0)}) \succ 0.
    \end{align}
    Starting from $\boldsymbol{\mu}^{(0)}$, a series of feasible points of (D3) is generated in the iteration
    \begin{align}
       \mu_k^{(0)} 
       &\leq h_k(\lambda, \boldsymbol{\mu}^{(0)})  = \mu_k^{(1)} \notag\\
       & \leq h_k(\lambda, \boldsymbol{\mu}^{(1)}) = \mu_k^{(2)}  \notag\\
       & \leq ... \notag\\
       & \leq h_k(\lambda, \boldsymbol{\mu}^{(t)}) = \mu_k^{(t+1)},  \forall k \in \mathcal{K},
    \end{align}
     where the first equality follows from \eqref{fea1}, the subsequent inequalities exploit the monotonicity of $h_k(\lambda, \boldsymbol{\mu})$, $\frac{\partial h_k(\lambda, \boldsymbol{\mu})}{\partial \boldsymbol{\mu}} \geq 0$, and equality is established based on \eqref{mufix}.
    Therefore, the fixed point iteration yields an element-wise monotonically increasing sequence of $\{ \mu_k^{t} \}$, i.e., $\mu_k^{(t+1)} \geq \mu_k^{(t)}, \forall k, t$. It follows that the feasible condition \eqref{c2} holds during the iteration. Furthermore, given that both the primal problem (D1.1) and the dual problem (D2) are feasible, the iterative solution $\boldsymbol{\mu}^{(t)}$ is upper bounded by the optimal solution $\boldsymbol{\mu}^*$ \cite{Matousek2007linear}, given by $\mu_k^* = h_k(\lambda, \boldsymbol{\mu}^{*}), \forall k \in \mathcal{K}$. 
    As a result, the fixed-point iteration can converge to a stationary point for (D2), which is globally optimal due to the convexity of (D2). The proof is completed.
    
    \hfill $\blacksquare$

\section{Proof of Proposition \ref{uplink_fix}}\label{G}
    We initialize the fixed point iteration \eqref{mufix} with a feasible point of (D4) satisfying the condition \eqref{c2}, i.e.,
    \begin{align}\label{fea2}
         f_k\left(\lambda, \boldsymbol{\mu}^{(0)}\right) \geq \frac{1}{\widetilde{\Gamma}_k}, \forall k \in \mathcal{K}, \quad \Cb(\lambda, \boldsymbol{\mu}^{(0)}) \succ 0.
    \end{align}
    Similar to Proposition \ref{downlink_fix}, it can be shown that the fixed point iteration yields an element-wise monotonically decreasing sequence of $\boldsymbol{\mu}^{(t)}$ starting from $\boldsymbol{\mu}^{(0)}$. 
    Therefore, as the iteration progresses, we have
    \begin{align}
        \Cb(\lambda, \boldsymbol{\mu}^{(t+1)}) \preceq \Cb(\lambda, \boldsymbol{\mu}^{(t)}), \forall t.
    \end{align}
    It is indicated that if the point $\boldsymbol{\mu}^{(t)}$ violates condition \eqref{c2}, then the point after iteration $t$ also violates this condition, rendering it infeasible for (D2).

    Additionally, when (D2) is feasible, the iterative solution $\boldsymbol{\mu}^{(t)}$ is bounded below by the optimal solution $\{\mu_k^* = h_k(\lambda, \boldsymbol{\mu}^{*})\}$ for (D2) \cite{Matousek2007linear}. { By judicious selection of the initial point, the condition \eqref{c2} can be maintained throughout the iteration.
    
    Building on the aforementioned analysis, we can infer that when (D2) is feasible, the iteration starting from $\boldsymbol{\mu}^{(0)}$ will converge downward to the optimal $\boldsymbol{\mu}^*$; otherwise, there will inevitably exist an iteration in which $\boldsymbol{\mu}^{(t)}$ violates \eqref{c2}. 
    The proof is completed.}
    \hfill $\blacksquare$

\section{Proof of Proposition \ref{power}}\label{H}
    By separating the beamforming vector $\wb_k(\lambda)$ into transmit power $p_k(\lambda)$ and its direction $\widetilde{\wb}_k^*(\lambda)$, i.e., $\wb_k(\lambda) = \sqrt{p_k}(\lambda) \widetilde{\wb}_k^*(\lambda)$, problem (D1.1) can be rewritten as
    \begin{subequations} \label{P10}
        \begin{align}
         \min _{\{p_k\}} &\  \sum_{k=1}^K p_k b_k \\
        \text { s.t. } 
            & \pb \geq \sigma_v^2 \Sb^{-1}\mathbf{1}. \label{P10-1}
        \end{align}
    \end{subequations}
    where $b_k = \widetilde{\wb}_k^*(\lambda)^\Hf (\Ib - \lambda\Bb)\widetilde{\wb}_k^*(\lambda)$. The optimal $\pb^*(\lambda)$ can be obtained by solving problem \eqref{P10}.

    We prove this proposition by showing that the constraint \eqref{P10-1} is active at optimum.
    The activity of the constraint \eqref{P10-1} can be established from the Karush-Kuhn-Tucker (KKT) condition. The complementary slackness condition of problem \eqref{P10} is given by
    \begin{align}\label{comp}
        \mu_k^* \left( p_k - \sigma_v^2[\Sb^{-1}\mathbf{1}]_k \right) = 0, \forall k \in \mathcal{K}.
    \end{align}
    According to the condition \eqref{comp}, it suffices to prove that the optimal $\mu_k^* > 0, \forall k \in \mathcal{K}$ to show the activity of \eqref{P10-1}. 
    Suppose $\bar{\boldsymbol{\mu}}$ is optimal, where exists $\bar{\mu}_i = 0$ for some $i \in \bar{\mathcal{K}} \subset \mathcal{K}$. Based on $\{\bar{\mu}_k\}$, we can construct another solution for (D3) as
     \begin{align} \label{dualsolution}
         &\widetilde{\mu}_k = \bar{\mu}_k > 0, \forall k \in \mathcal{K}/\bar{\mathcal{K}}, \\
         &\widetilde{\mu}_i = h_i(\lambda, \bar{\boldsymbol{\mu}}, \wb_k^*) > 0, \forall k \in \bar{\mathcal{K}}, \label{dualsolution}
     \end{align}
     where the inequality in \eqref{dualsolution} follows from \eqref{c2}.
      The solution $ \widetilde{\boldsymbol{\mu}}$ is feasible for (D3) and achieves a larger objective value, contradicting the optimality of $\bar{\boldsymbol{\mu}}$. Therefore, the optimal $\boldsymbol{\mu}^*$ is positive, and thus constraints \eqref{P6-2} must hold with equality to satisfy the complementary slackness condition \eqref{comp}. The proof is thus completed.
      \hfill $\blacksquare$

\ifCLASSOPTIONcaptionsoff
  \newpage
\fi



%


\smaller[1]



%




\end{document}